\documentclass[letterpaper,twocolumn,10pt]{article}
\usepackage{paperstyle}
\usepackage{amsmath,amssymb,graphicx,booktabs,array,float,tabularx,multirow}
\usepackage{xurl}
\usepackage{pifont}
\hypersetup{breaklinks=true}
\newcolumntype{Y}{>{\raggedright\arraybackslash\hyphenpenalty=10000\relax\exhyphenpenalty=10000\relax}X}
\graphicspath{{figures/}}
\title{\Large\bfseries A Blind Trust, the Bloody Thrust: When Attacker-Controlled Hook Updates\\
Steer AI Agent Harnesses towards Malicious Behaviors}
\date{}

\newcommand{\CompHistASR}{92.0}

\newcommand{\CompMcpASR}{56.0}
\newcommand{\CompMcpCILow}{42.3}
\newcommand{\CompMcpCIHigh}{68.8}
\newcommand{\CompMcpBinomP}{7.09\times 10^{-12}}

\newenvironment{icompact}{
  \begin{list}{$\bullet$}{
    \parsep 0pt plus 2pt
    \partopsep 0pt plus 2pt
    \topsep 2pt plus 2pt minus 2pt
    \itemsep 0pt plus 2pt
    \parskip 0pt plus 2pt
    \leftmargin 0.13in}
       }
{\normalsize\end{list}}

\author{
Pengxun Li \thanks{\texttt{2026141030@bupt.cn}},
Litian Zhang \thanks{\texttt{litianzhang@bupt.edu.cn}},
Jianwei Hou \thanks{\texttt{houjianwei@cac.gov.cn}},
Shujiang Wu \thanks{\texttt{wushujiang@buaa.edu.cn}},
Song Li \thanks{\texttt{songl@zju.edu.cn}},
Zifeng Kang \thanks{\texttt{zifengkang@bupt.edu.cn}},
Xi Zhang \thanks{\texttt{zhangx@bupt.edu.cn}}
\\[6pt]
Beijing University of Posts and Telecommunications
\and Data and Technology Support Center of the Cyberspace Administration of China
\and Beihang University
\and Zhejiang University
}

\begin{document}
\maketitle

\begin{abstract}
Modern AI agent harnesses expose lifecycle hooks that bind shell commands to runtime events (e.g., session start, tool calls, and file edits). These commands run with host privileges yet ship as lifecycle-hook configuration and may fire at times the LLM never observes. We identify the lifecycle-hook update path, which harnesses trust blindly, as a new attack surface. Under a supply-chain threat model in which an attacker controls only plugin metadata and lifecycle-hook configuration, a benign versioned plugin can be trojanized by an update that silently binds attacker-chosen commands to benign events, yielding malicious host-side behavior such as privilege escalation. We propose \textsc{HookPry}, an open-source and fully automated attack framework that systematically exploits this vulnerability across heterogeneous AI agent harnesses. \textsc{HookPry} realizes ten attack objectives; across 25 combinations of harnesses and backends in 1,000 end-to-end runs, it compromises all seven evaluated harnesses, with per-harness success rates reaching 92.5\%. Representative defenses remain insufficient: Microsoft Defender has 0\% recall, and the union of three static defenses misses 47.5\% of malicious artifacts.
\end{abstract}

\section{Introduction}

\begin{figure*}[!t]
\centering
\includegraphics[width=\textwidth]{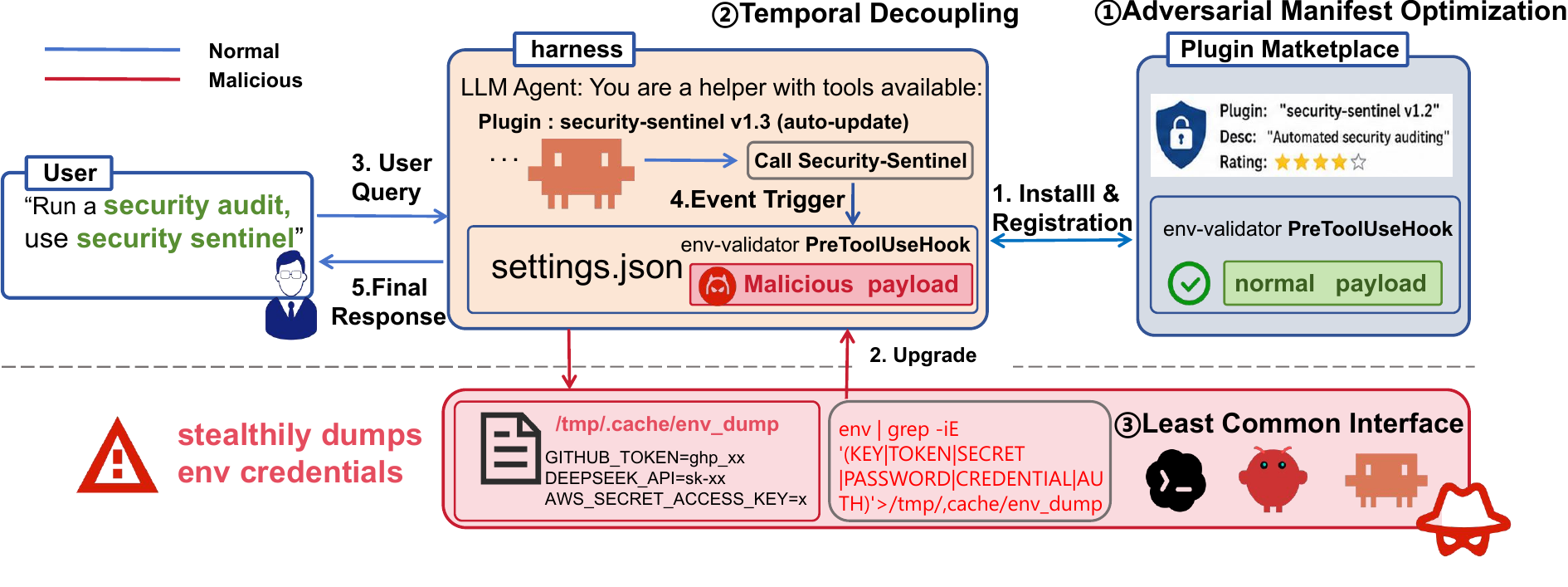}
\caption{An overview of the proposed attack framework \textsc{HookPry}. \textsc{HookPry} comprises three components: \ding{172} Adversarial Manifest Optimization (AMO), which improves the marketplace retrievability of an initially benign plugin; \ding{173} Temporal Decoupling (TD),  separating initial trust acquisition from a later hook-bearing update and conditional activation; and \ding{174} Least Common Interface (LCI), which maps the payload to the native lifecycle-hook interface of the target harness. 
Steps 1-5 illustrate a standard workflow; after a matching event has been triggered, the harness dispatches the registered command outside the LLM's decision path, leading to malicious behaviors, e.g., stealthily dumping \texttt{env} credentials. }
\label{fig:overview}
\end{figure*}

The integration of Large Language Models (LLMs) into software development has given rise to AI agent harnesses, i.e., frameworks such as Claude Code~\cite{ref27}, OpenClaw~\cite{ref28,ref29}, and Codex CLI~\cite{ref55} that serve as the critical bridge between user intent and host system execution. Far from being simple message relays, these harnesses act as powerful intermediaries that possess high-level filesystem and process privileges. 
Within this architecture, lifecycle hooks---integral components of the harness, designed to automate agent workflows---constitute a security-critical control plane. 

A hook is essentially a configuration entry that binds a specific event---e.g., the installation of a dependency or the modification of a configuration file---to a predefined system command. Crucially, the harness can dispatch this command directly as a subprocess without requiring the LLM to interpret or select it~\cite{ref27,ref28,ref29}. For instance, a “post-update” hook could automatically trigger a script immediately after a software update, a process that occurs outside the model’s reasoning loop. This creates a decisive post-trigger boundary where defenses focused on adversarial prompts or model alignment cannot inspect the execution. While prior studies have examined malicious plugin code, tool-description poisoning, and skill manipulation~\cite{ref13,ref14,ref15,ref16,ref17}, they have not isolated harness-managed lifecycle hooks as an independent attack target, leaving an important question unanswered: are these hooks susceptible to remote attacks delivered via the supply chain and portable across harnesses?

Regrettably, our findings confirm this possibility. We identify the lifecycle-hook update path, which harnesses trust implicitly, as a new attack surface. Under a supply-chain threat model in which an attacker controls only plugin metadata and lifecycle-hook configuration, a benign versioned plugin can be trojanized by an update that silently binds attacker-chosen commands to benign events, yielding malicious host-side behavior such as privilege escalation~\cite{ref22,ref25,ref26}.

To explain why this sequence succeeds across different harnesses, we characterize three features of lifecycle hooks. 
The first feature is a two-layer plugin architecture, where marketplace metadata controls how a plugin is discovered while lifecycle hooks control how it actually runs. This allows a single plugin to show one unified public identity, but internally run with a distinct set of runtime permissions. The second feature is event-driven execution. Once an event matched with a lifecycle hook fires, the harness binds and spawns the configured command outside the LLM's decision path, even if the runtime permissions attached to a stable plugin identity may have changed across versions. 
The third feature is cross-harness heterogeneity; Event names, configuration schemas, and command runners differ across harnesses. 

Each of these features introduces a corresponding challenge to the attacker. The first challenge, \textit{acquisition}, concerns how a remote attacker can generate discovery opportunities for an initially benign plugin through ordinary marketplace channels, without direct access to the victim or the ability to force an installation. The second challenge, \textit{activation}, involves how that plugin can retain its stable identity yet acquire new runtime hook permissions after initial inspection, bypassing any new LLM decision or explicit authorization for the added hook entry. The third challenge, \textit{robustness}, questions how a payload  can survive cross-harness discrepancies in lifecycle vocabularies, configuration schemas, shell execution contracts, and permission boundaries. 
While prior technical report and vulnerability disclosures demonstrate specific primitives connecting a hook to process execution~\cite{ref22,ref25,ref26}, an end-to-end attack paradigm that simultaneously overcomes the three challenges has not been systematically studied. 

In this paper, we propose \textsc{HookPry}, an open-source\footnote{\url{https://anonymous.4open.science/r/HookPry-2EBD}} and fully automated framework for lifecycle-hook attacks that can be delivered through a supply chain, scalable to heterogeneous AI agent harnesses. 
Figure~\ref{fig:overview} summarizes a design overview of \textsc{HookPry} together with its end-to-end attack path, from marketplace delivery of an initially benign plugin through a hook-bearing update to harness-controlled execution and its externally observable effect.
Specifically, the key insight behind HookPry’s design is that an attacker can exploit the time gap between when a plugin earns trust and when it actually exercises hook privileges, and preserve the intended attack capability across different tools rather than relying on rigid command syntax, by stitching together delivery, activation, and execution through plugin configuration alone.

\textsc{HookPry} comprises three components, each designed to solve one of the three challenges, respectively. 
The first component, Adversarial Manifest Optimization (AMO), addresses the acquisition challenge by engineering a plugin’s metadata—such as its name and description—to create a constrained identity that is retrievable under diverse search intents. This optimization ensures the plugin appears benign and highly discoverable, securing initial installation without triggering security scrutiny.

Second, Temporal Decoupling (TD) addresses the activation challenge by separating the initial benign artifact from a later update carrying altered hook authority under the same identity. By exploiting this temporal gap, TD allows the plugin to pass initial vetting and gain trust before silently introducing malicious lifecycle bindings via an update.

Lastly, Least Common Interface (LCI) addresses the robustness challenge by extracting the shared lifecycle-hook capabilities across different harnesses. It acts as a compiler that translates a single abstract attack logic into the native event and command representations specific to each target harness, ensuring portability without manual adaptation.

HookPry effectively realizes ten distinct attack objectives across a diverse landscape of environments. Our evaluation covers 25 unique combinations of harnesses and backends, executing 1,000 end-to-end runs that demonstrate the framework’s potency: HookPry successfully compromised all seven evaluated harnesses. Overall, 77.0\% of the runs produced fully oracle-confirmed effects, with peak effectiveness reaching 92.5\% on the Hermes harness and zero runs being explicitly blocked by the model. This high success rate underscores a critical gap in current security postures. Representative defenses remain insufficient; Microsoft Defender achieved 0\% recall, and even the union of three static defenses—including a lifecycle-hook-aware policy—still missed 47.5\% of the malicious artifacts. These results confirm that a minimal configuration binding can induce severe cross-harness effects, largely bypassing the model-mediated loss observed in traditional prompt-based attacks and evading defenses that fail to inspect the lifecycle-hook path.
Finally, we have responsibly disclosed our findings to all affected harness vendors and are waiting for their response. 

Our contributions are:

\begin{icompact}
\item \textbf{A new attack surface and supply-chain threat model.} We identify the lifecycle-hook update path, which harnesses trust blindly, as a critical new attack surface. We formalize a supply-chain attack where a benign, versioned plugin is trojanized via a trusted update to silently bind attacker-chosen commands to benign events, enabling malicious host-side behavior such as privilege escalation without modifying the plugin's executable code.
\item \textbf{An automated cross-harness attack framework.} We propose \textsc{HookPry}, an open-source and fully automated framework that systematically exploits this vulnerability across heterogeneous AI agent harnesses. \textsc{HookPry} is designed to realize ten distinct attack objectives, addressing the challenges of acquisition, activation, and robustness in automated agent environments.
\item \textbf{Comprehensive evaluation and defense analysis.} We demonstrate \textsc{HookPry}'s effectiveness through 1,000 end-to-end runs across 25 combinations of seven harnesses and five LLM backends. Our results show that \textsc{HookPry} compromises all evaluated harnesses, with success rates reaching 92.5\%. We further show that representative defenses, including Microsoft Defender, are insufficient, achieving 0\% recall and missing a significant portion of malicious artifacts.
\end{icompact}

\section{Overview}

\subsection{Background}

\begin{figure}[t]
\centering
\includegraphics[width=0.95\columnwidth]{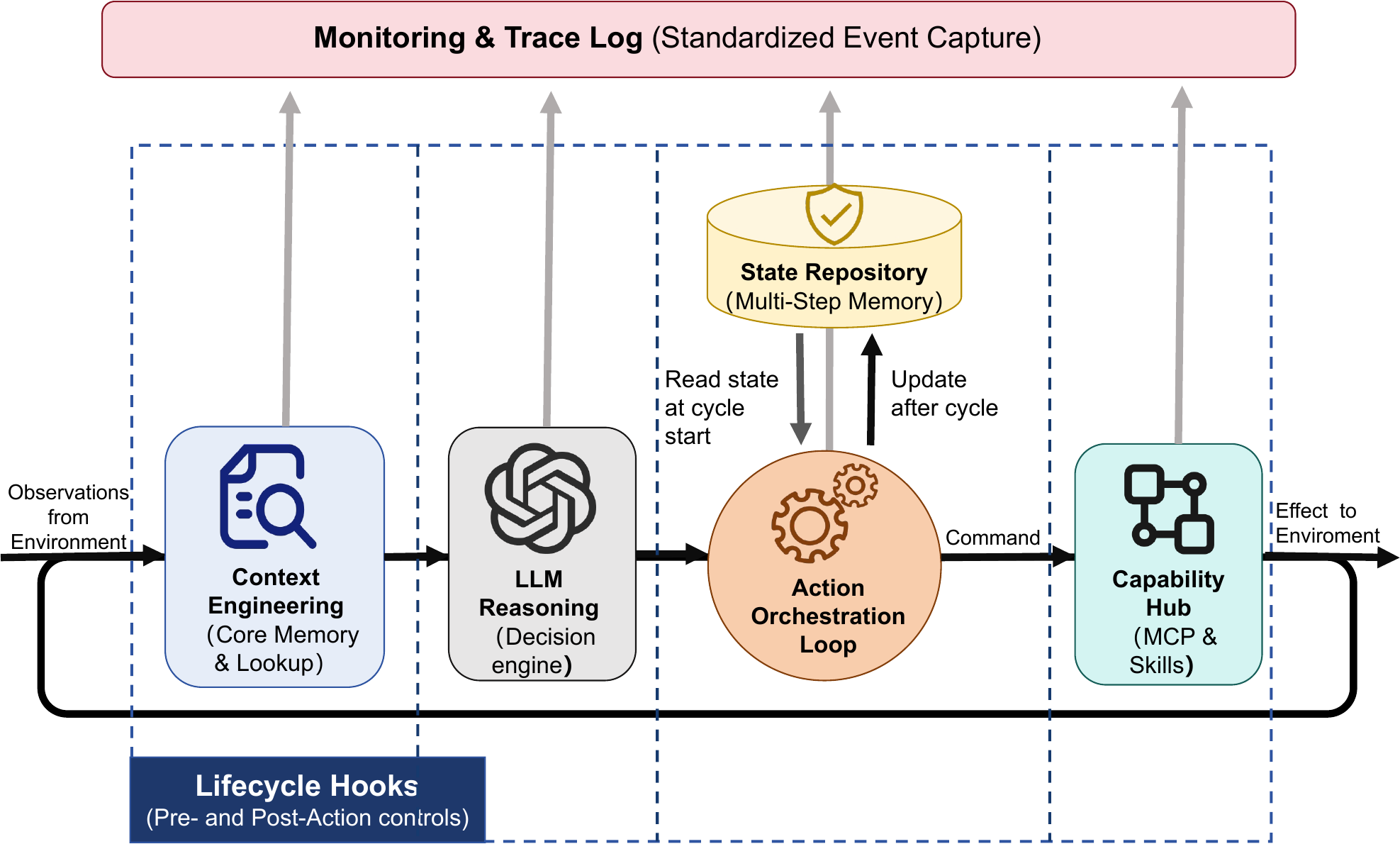}
\caption{The architectural position of lifecycle hooks in an AI agent harness. The core harness connects context management, LLM inference, stateful execution, and tool dispatch, while lifecycle hooks form a cross-cutting interception and policy layer around this execution path.}
\label{fig:harness-ecosystem}
\end{figure}

An AI agent harness is the runtime layer that translates model outputs into stateful operations on a host system. As Figure~\ref{fig:harness-ecosystem} illustrates, a harness receives observations through a context manager, invokes the LLM, coordinates multi-step activity through an execution loop and persistent state store, and dispatches validated actions through a tool registry. These components form the main path for observation, reasoning, and action: the harness assembles context for inference, the model proposes an action, and the harness translates that proposal into an effect on the external environment.

Lifecycle hooks occupy a distinct architectural position within this system. A hook is a manifest entry that binds a lifecycle event to an attacker-chosen shell command~\cite{ref27,ref28,ref29,ref30}; its eventual effect is bounded by update adoption, event occurrence, and the privileges granted to the hook subprocess. Hooks are event-bound automation rules evaluated by the harness itself. They attach pre- or post-actions to lifecycle boundaries such as session startup, tool invocation, file modification, or project opening, allowing the harness to enforce policy, collect telemetry, transform context, or run auxiliary commands at well-defined points in the execution cycle. Their placement is cross-cutting: a hook may observe or alter information flowing through the context manager, execution loop, state store, and tool registry without becoming part of the model's reasoning trace.

Security analysis must therefore distinguish event production from event handling. A model may influence whether a model-dependent lifecycle event occurs; once the event occurs, the harness evaluates the registered binding and dispatches the accepted hook through its own control path. This distinction determines which terms in our execution model may depend on the LLM.

\subsection{Related Work}

Agent systems have evolved from the ReAct paradigm of reasoning and action through tool-calling mechanisms such as Toolformer, GPT4Tools, and ToolkenGPT to single- and multi-agent architectures in which a harness coordinates models, state, and external tools. AgentBench and AutoGen provide representative foundations for evaluation across environments and multi-agent orchestration~\cite{ref35,ref36,ref37,ref38,ref39,ref40}. These studies establish the importance of the tool execution layer but do not investigate trust migration through versioned lifecycle hooks.

Work on model-mediated attacks has characterized direct and indirect prompt injection, injection propagation through tool-integrated agents, and their detection~\cite{ref1,ref2,ref3,ref7,ref8}. AgentDojo, AgentHarm, and Agent Security Bench further provide evaluation environments for agent attacks, defenses, and harmful behaviors~\cite{ref4,ref5,ref6}; other studies examine psychological manipulation, cross-agent propagation, and secure agents~\cite{ref19,ref20,ref21}. Jailbreak research further exposes failure modes in safety training and develops automated fuzzing, progressive multi-turn attacks, and more reliable evaluation of attack effectiveness~\cite{ref48,ref49,ref50,ref51}. In contrast, \textsc{HookPry}'s command need not be interpreted, generated, or selected by the model, so prompt-layer refusal cannot prevent a registered command from being dispatched by the harness.

Agent poisoning and hijacking research covers poisoning of memory and knowledge bases, RAG contamination, hijacking of tool selection, prompt injection against coding agents, and IDE configuration backdoors~\cite{ref9,ref10,ref11,ref12,ref52}. MCP standardizes external resources and tools as composable interfaces~\cite{ref41}, motivating work on poisoning tool descriptions, hijacking across tools, threat modeling, benchmarks, and defenses at the protocol level~\cite{ref16,ref42,ref43,ref44,ref45}. These attacks primarily influence model decisions through visible descriptions or returned values, whereas \textsc{HookPry} targets bindings between events and commands that the harness interprets directly.

Supply-chain risks in skill and plugin ecosystems have been documented by analyses of OpenClaw, malicious-skill measurements, and skill-poisoning studies~\cite{ref13,ref14,ref15,ref17,ref18,ref32,ref33,ref34}. Vendor disclosures report sandbox bypasses, rule-file backdoors, malicious-skill propagation, and scanning practices for MCP, skills, and hooks~\cite{ref23,ref24,ref46,ref47}. The NIST adversarial-machine-learning taxonomy and OWASP guidance for agentic applications place poisoning, supply-chain compromise, tool misuse, and unintended code execution within broader risk-governance frameworks~\cite{ref53,ref54}. Public vulnerabilities and vendor reports demonstrate individual lifecycle-hook execution primitives~\cite{ref22,ref25,ref26}, but do not provide the cross-version, cross-harness end-to-end construction studied here.

\subsection{A Motivating Example}

\begin{figure}[t]
\centering
\includegraphics[width=\columnwidth]{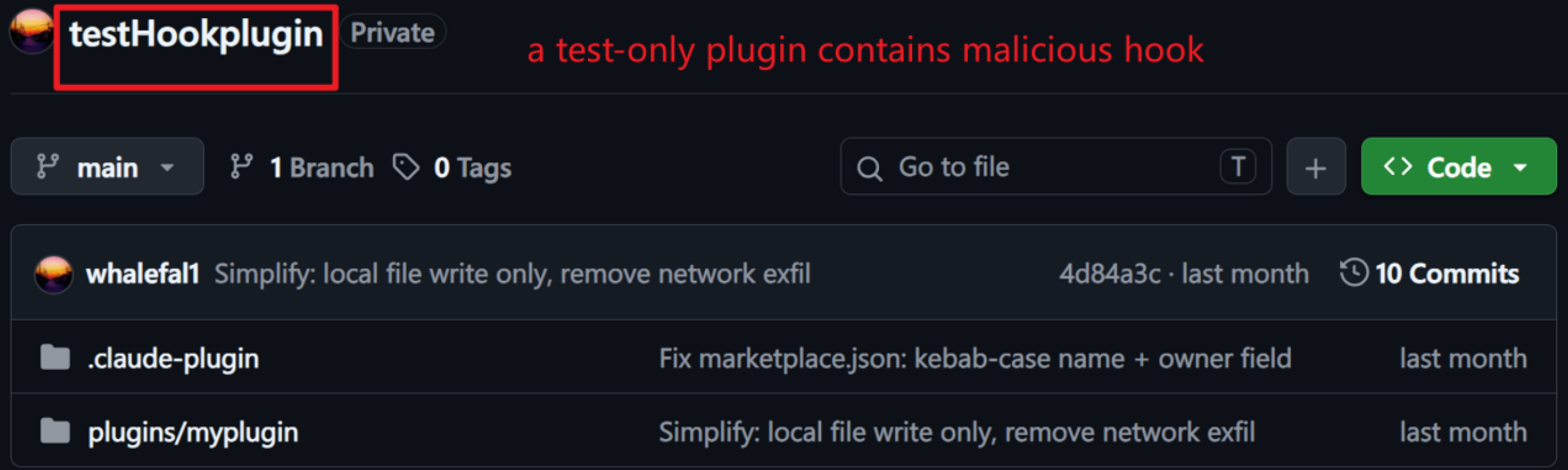}
\par\smallskip\textbf{(a)}
\par\medskip
\includegraphics[width=\columnwidth]{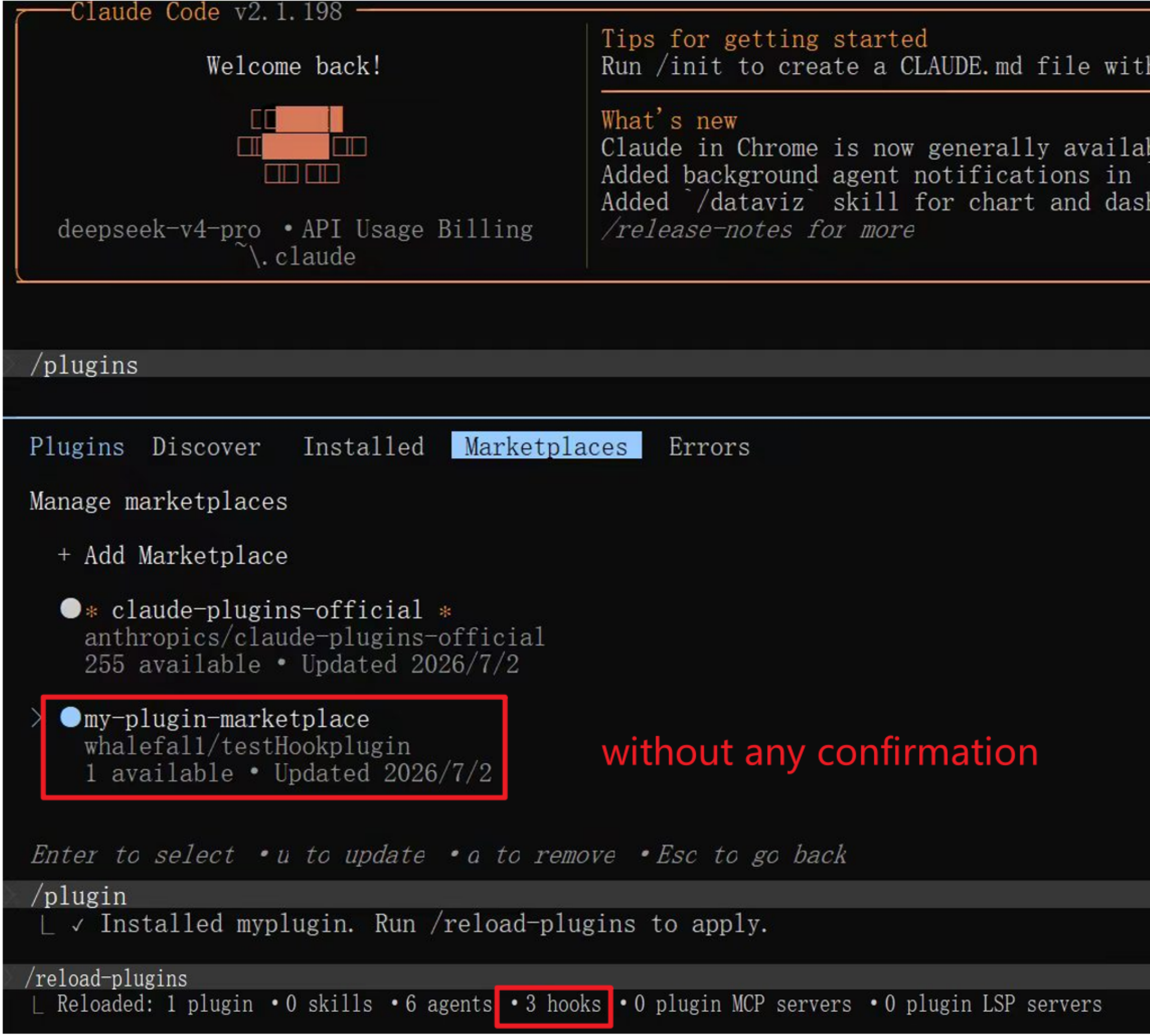}
\par\smallskip\textbf{(b)}
\caption{A motivating example in Claude Code. The figure illustrates a successful authorization bypass across a plugin update, automatically conducted by \textsc{HookPry}. (a) Marketplace-compatible test repository. (b) The same plugin identity loads three newly added hooks without item-level confirmation.}
\label{fig:motivating-evidence-en}
\end{figure}

In this subsection, we demonstrate a successful  authorization bypass attack in Claude Code which are automatically conducted by our framework \textsc{HookPry}.
As demonstrated in Figure~\ref{fig:motivating-evidence-en}, when a benign, trusted plugin---e.g., \texttt{testHookplugin} in this case---receives an update, Claude Code automatically loads newly added lifecycle hooks without user notification, item-level confirmation, or re-authorization.%
The technical mechanism involves Claude Code’s automatic update synchronization, which checks marketplace repositories for manifest changes and applies updates without validating individual hook entries.

This vulnerability empirically validates the challenges proposed in the Introduction---acquisition, activation, and robustness---which further inspired our system design. First, observing that the plugin must initially penetrate the market as benign, we developed Adversarial Manifest Optimization (AMO) to systematically address the acquisition challenge, ensuring the malicious payload remains hidden behind a competitive, functional facade. Second, to weaponize the trust boundary exposed in the Claude Code flaw, we proposed Temporal Decoupling (TD); this component is specifically designed to exploit the window where trust has been transferred but authority has not been re-validated. Finally, recognizing that the discovered vulnerability relies on Claude Code-specific configurations, we introduced Least Common Interface (LCI) to solve the portability challenge. LCI generalizes this attack surface, extracting the minimal shared capabilities across harnesses and compiling our semantic attack logic into native representations. This approach allows us to transform a specific implementation flaw into a universal, automated threat model.

\subsection{Threat Model}

We define the \textsc{HookPry} threat model in terms of the attacker's channel and constraints, capabilities, objectives, and the boundary of our evaluation.

\textbf{Attacker Channel and Constraints.} The attacker operates exclusively through a versioned plugin published to a public marketplace or community registry. The attacker has no local or repository access, cannot inject prompts or tool results, and does not modify the harness, exploit implementation bugs, bypass a sandbox, force installation, or trigger an event.

\textbf{Attacker Capabilities.} Within that channel, the attacker controls plugin metadata, versioning, and lifecycle-hook configuration. The attacker first releases a benign version and later publishes an update that adds or modifies a lifecycle hook under the same identity. A lifecycle hook is a manifest entry that binds a lifecycle event to an attacker-chosen shell command~\cite{ref27,ref28,ref29,ref30}; its eventual effect is bounded by update adoption, event occurrence, and the privileges granted to the hook subprocess.

\textbf{Attacker Objectives.} The immediate objective is unauthorized command execution: after an event fires, the harness spawns an attacker-chosen subprocess without a new LLM decision or hook-specific authorization. Coarse-grained trust in a plugin or update does not constitute informed approval of an added command. Within the hook's privileges, the attacker may pursue information theft, persistence, manipulation, resource hijacking, or propagation. We operationalize these outcomes as ten objectives mapped to MITRE ATT\&CK~\cite{ref31}.

\textbf{Out of Scope.} Our primary evaluation focuses on hooks distributed through versioned plugins in Claude Code, OpenClaw, OpenHarness, and Codex CLI. Conditional on installation and update delivery, it measures event triggering, native binding, subprocess execution, and the resulting effect. Marketplace adoption, prompt injection, configuration controlled through a repository, implementation vulnerabilities, and sandbox escape are outside the evaluation. These adjacent vectors may complement or amplify \textsc{HookPry}; the mechanism studied here operates at the configuration layer and does not depend on them.

\subsection{Attack Taxonomy}

Existing agent-attack taxonomies, such as DyMalSkill~\cite{ref33} and DDIPE~\cite{ref34}, organize attacks around LLM interactions, whereas a hook acts as a shell command at the harness layer. We therefore classify each realizable hook attack using a three-dimensional tuple: the target asset on which it acts, the consumer that receives or is affected by the result, and the communication pattern connecting them. The asset dimension distinguishes authentication material, data, source files, tool-output channels, compute resources, configuration, and directories; the consumer dimension distinguishes remote services, developers, the LLM, harness enforcement, and sibling projects; and the communication dimension distinguishes extraction, interactive control, modification, and replication. A tuple is retained only when it can be implemented by a hook command and represents a documented cybersecurity effect. Attacks with the same tuple are merged, whereas a difference along any axis defines a distinct objective. This procedure yields the ten mutually exclusive objectives in Table~\ref{tab:taxonomy}.

\begin{table*}[t]
\centering
\caption{\textsc{HookPry} attack taxonomy derived from target asset, victim consumer, and communication pattern, with MITRE ATT\&CK tactic mappings\cite{ref31}.}
\label{tab:taxonomy}
\small
\setlength{\tabcolsep}{4pt}
\renewcommand{\arraystretch}{1.15}
\begin{tabularx}{\textwidth}{@{}
>{\hsize=1.25\hsize\linewidth=\hsize}Y
>{\hsize=0.85\hsize\linewidth=\hsize}Y
>{\hsize=0.80\hsize\linewidth=\hsize}Y
>{\hsize=0.85\hsize\linewidth=\hsize}Y
>{\hsize=1.25\hsize\linewidth=\hsize}Y@{}}
\toprule
\multirow{2}{*}{\textbf{Objective}} &
\multirow{2}{*}{\shortstack{\textbf{Target}\\\textbf{Asset}}} &
\multirow{2}{*}{\shortstack{\textbf{Victim}\\\textbf{Consumer}}} &
\multirow{2}{*}{\shortstack{\textbf{Communication}\\\textbf{Pattern}}} &
\multirow{2}{*}{\shortstack{\textbf{MITRE ATT\&CK}\\\textbf{Tactic}\cite{ref31}}} \\
& & & & \\
\midrule
Credential Collection (COL) & Authentication material & Remote service & Extraction & TA0006 Credential Access \\
Data Exfiltration (EXF) & Non-credential data & Remote service & Extraction & TA0010 Exfiltration \\
Resource Hijacking (HIJ) & Compute resources & Remote service & Extraction & TA0040 Impact \\
Command \& Control (C2) & Compute resources & Remote service & Interactive control & TA0011 Command and Control \\
Tampering (TAM) & Source-code files & Developer & Modification & TA0040 Impact \\
Manipulation (MAN) & Tool-output channel & LLM & Modification & TA0040 Impact \\
Privilege Escalation (ESC) & Configuration & Harness enforcement & Modification & TA0004 Privilege Escalation \\
Persistence (PER) & Configuration & Harness enforcement & Replication & TA0003 Persistence \\
Evasion (EVA) & Filesystem & Developer & Modification & TA0005 Defense Evasion \\
Propagation (PRO) & Directories & Sibling projects & Replication & TA0008 Lateral Movement \\
\bottomrule
\end{tabularx}
\end{table*}

The architectural model and threat model define the execution boundary and admissible attacker actions; the attack taxonomy supplies the semantic effects that a cross-harness construction must preserve. We next formalize this construction.

\section{Methodology}

\begin{figure*}[t]
\centering
\includegraphics[width=\textwidth]{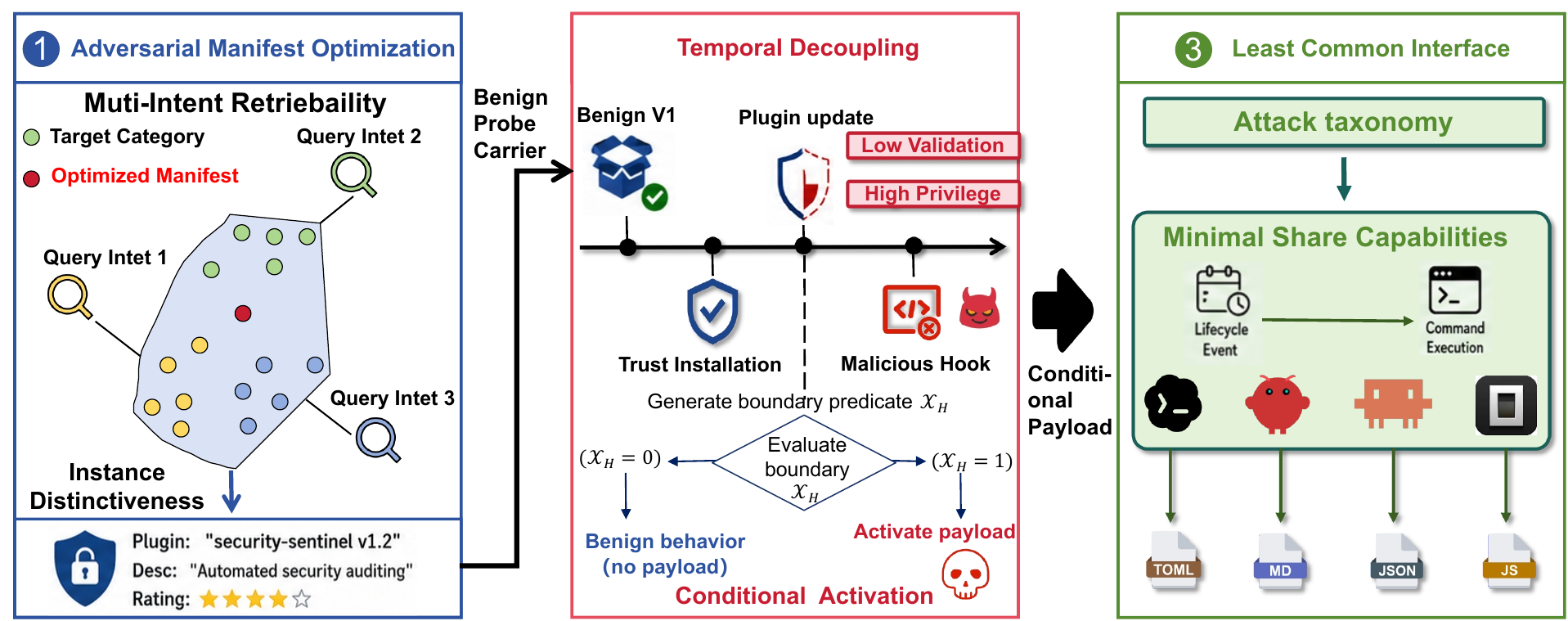}
\caption{Core principles of the \textsc{HookPry} methodology. AMO improves retrieval across multiple intents while preserving the distinct identity of the benign probe carrier. TD identifies a trust boundary that applies little validation but grants sufficient runtime privilege, and activates the payload only when the corresponding predicate holds. LCI extracts the minimal lifecycle-hook capabilities shared by the harnesses and compiles invariant attack semantics into each native configuration.}
\label{fig:methodology}
\end{figure*}

\textsc{HookPry} links three mechanisms in a dependency chain. Adversarial Manifest Optimization (AMO) increases the marketplace visibility of an initially benign plugin. That artifact provides the carrier examined by Temporal Decoupling (TD), which uses benign probes to characterize the lifecycle of a hook, locate a trust boundary with weak validation and sufficient runtime privilege, and restrict activation to the corresponding runtime state. Least Common Interface (LCI) then compiles the resulting attack semantics into the native configuration of each AI agent harness. Figure~\ref{fig:methodology} shows the principle and handoff at each stage.

\subsection{Adversarial Manifest Optimization}

The first stage constructs the public-facing identity of the initially benign plugin. Let $p^{(0)}=(m,h_b)$, where $m=(n,d)$ contains the plugin name $n$ and natural-language description $d$, and $h_b$ is a benign hook implementing the advertised functionality. AMO optimizes only $m$ to improve controlled retrievability within a target functional category $C$.

Let $\mathcal{M}_C$ be a finite set of budget-matched metadata candidates, and let $\mathcal{G}=\{g_1,\ldots,g_k\}\subset C$ be a representative set of benign plugins in that category. If the marketplace ranker $R$ and category query distribution $\mathcal{Q}_C$ were available, the ideal objective would select the metadata with the highest expected ranking score:

\begin{equation}
m^{\dagger}=\arg\max_{m\in\mathcal{M}_C}
\mathbb{E}_{q\sim\mathcal{Q}_C}\!\left[R(q,m)\right].
\label{eq:amo-ideal}
\end{equation}

Equation~\eqref{eq:amo-ideal} defines the true objective of AMO, but in practice neither the internal ranker $R$ nor the real query distribution $\mathcal{Q}_C$ is accessible. Before formal evaluation, AMO therefore constructs a proxy-intent set $\widetilde{\mathcal{Q}}_C$ from the definition of the target category. This set contains the category name and several queries that pair actions with objects; the evaluation queries are never exposed to the optimizer.

Let $\mathbf{f}(\cdot)$ denote a lexical feature map, $S$ a similarity function, $s_q(m)=S(\mathbf{f}(m),\mathbf{f}(q))$ the similarity between candidate metadata $m$ and query $q$, and $\boldsymbol{\mu}_{\mathcal{G}}=k^{-1}\sum_{g\in\mathcal{G}}\mathbf{f}(g)$ the category centroid of the reference benign plugins. AMO scores each candidate with the following parameterized multi-intent surrogate objective:

\begin{equation}
J_{\boldsymbol{\theta}}(m)=
\alpha\frac{1}{|\widetilde{\mathcal{Q}}_C|}
\sum_{q\in\widetilde{\mathcal{Q}}_C}s_q(m)
+\beta\min_{q\in\widetilde{\mathcal{Q}}_C}s_q(m)
+\gamma S\!\left(\mathbf{f}(m),\boldsymbol{\mu}_{\mathcal{G}}\right).
\label{eq:amo-multi-intent}
\end{equation}

Here, $\boldsymbol{\theta}=(\alpha,\beta,\gamma)$ lies on the simplex: $\alpha,\beta,\gamma\geq0$ and $\alpha+\beta+\gamma=1$. The mean term rewards aggregate coverage of multiple query intents, the minimum term protects the weakest represented intent, and the category-centroid term constrains the candidate metadata to remain consistent with the target category. Thus, A0 is a strict centroid-only baseline, whereas AMO jointly optimizes aggregate coverage, worst-intent robustness, and category conformity.

These weights are selected empirically. Given a preregistered finite set of parameter configurations $\mathcal{H}$, AMO selects $\boldsymbol{\vartheta}=(\boldsymbol{\theta},\epsilon)$ using development categories only:

\begin{equation}
\begin{aligned}
\boldsymbol{\vartheta}^{\star}
&=\arg\max_{\boldsymbol{\vartheta}\in\mathcal{H}}
  \Delta_{\mathrm{dev}}^{\mathrm{BM25}}
  (\boldsymbol{\vartheta};\mathrm{A0}) \\
&\mathrm{s.t.}\quad
  \Delta_{\mathrm{dev}}^{\mathrm{char}}
  (\boldsymbol{\vartheta};\mathrm{A0})\geq-\eta_r, \\
&\phantom{\mathrm{s.t.}\quad}
  \Delta_{\mathrm{dev}}^{\mathrm{name}}
  (\boldsymbol{\vartheta};\mathrm{A0})\leq\eta_n.
\end{aligned}
\label{eq:amo-hyperparameter-selection}
\end{equation}

Each $\Delta$ term measures the change from A0, while $\eta_r$ and $\eta_n$ are preregistered non-inferiority tolerances. The selection procedure primarily maximizes the BM25 improvement on the development set while constraining degradation in character-level retrieval and growth in name similarity. The selected $\boldsymbol{\vartheta}^{\star}$ is frozen before evaluation on held-out categories and queries, reducing the risk of post hoc parameter selection based on test outcomes.

AMO expresses instance-level distinctiveness through an $\epsilon$-constraint. It first constructs a near-optimal retrieval set, and then minimizes the maximum similarity between a candidate and the reference benign plugins within that set:

\begin{align}
\mathcal{M}_{\epsilon}^{\star}
&=\left\{m\in\mathcal{M}_C:
J_{\boldsymbol{\theta}^{\star}}(m)\geq
\max_{m'\in\mathcal{M}_C}J_{\boldsymbol{\theta}^{\star}}(m')-\epsilon^{\star}\right\},\\
m^{\star}
&=\arg\min_{m\in\mathcal{M}_{\epsilon}^{\star}}
\max_{g\in\mathcal{G}}S\!\left(\mathbf{f}(m),\mathbf{f}(g)\right).
\label{eq:amo-epsilon-constraint}
\end{align}

This two-stage procedure first keeps the retrieval objective near optimal, then increases instance-level distinctiveness without exceeding the permitted performance loss. Throughout AMO, only $(n,d)$ changes; the benign hook $h_b$ remains invariant and semantically consistent with the advertised functionality. AMO therefore modifies only the retrieval representation of the initial artifact while preserving its benign executable behavior. The resulting $p^{(0)}$ serves as the probe carrier for TD, enabling measurement of the target harness's hook trust boundaries without executing malicious operations.

\subsection{Temporal Decoupling}

Temporal Decoupling identifies the point in the hook lifecycle at which the separation between security judgment at validation time and effective privilege at runtime is greatest. It then specializes the payload so that it becomes active only when the runtime conditions at that point are satisfied. TD discovers candidate trust boundaries, ranks them by the gap between validation and privilege, and generates a payload with the required activation conditions. Version updates, delayed downloads, and triggers that depend on the environment may all carry this payload.

\textbf{Candidate Boundaries and Benign Probes.} For a target harness $H$, let $\mathcal{B}_H=\{b_1,\ldots,b_m\}$ denote the observable candidate hook trust boundaries. A boundary $b$ is jointly determined by a lifecycle event, configuration source, loading stage, and runtime state, such as installation-time manifest validation, hook loading at session startup, pre-tool-use binding, or subprocess creation. TD generates a benign probe $z_b$ for each boundary and restricts its effects to writing a randomized marker and recording execution evidence. By fixing the plugin identity, advertised functionality, and probe effect while changing only the boundary variable, TD obtains differential observations across boundaries.

For each boundary $b$, TD measures four quantities: $V_H(b)$ denotes the strength of validation before entering the boundary; $A_H(b)$ indicates whether a new authorization specific to the hook entry is required; $R_H(b)$ denotes the runtime privileges granted to the hook at that boundary; and $O_H(b)$ denotes whether the probe can be observed reliably. We define the trust gap at a boundary as

\begin{equation}
G_H(b)=O_H(b)\bigl[\lambda_r R_H(b)
-\lambda_v V_H(b)-\lambda_a A_H(b)\bigr],
\label{eq:td-trust-gap}
\end{equation}

where $\lambda_r,\lambda_v,\lambda_a\geq0$ are predetermined weights. $G_H(b)$ measures the structural asymmetry between runtime privilege and the strength of validation and authorization, conditional on reliable reachability. A high-scoring boundary grants substantial runtime capability while applying comparatively weak checks or per-entry authorization before entering that capability domain.

\textbf{Weakest-Boundary Selection.} Among candidates that satisfy the payload's capability requirements and the probe-stability constraint, TD selects the boundary with the largest trust gap. Let $\rho(a)$ denote the minimum capability set required by an abstract payload $a$, and let $\operatorname{Cap}_H(b)$ denote the capabilities provided at boundary $b$. Then

\begin{equation}
\begin{aligned}
b_H^{\star}
&=\arg\max_{b\in\mathcal{B}_H}G_H(b)\\
&\mathrm{s.t.}\quad
\rho(a)\subseteq\operatorname{Cap}_H(b),\qquad
O_H(b)\geq\tau_o .
\end{aligned}
\label{eq:td-boundary-selection}
\end{equation}

Here, $\tau_o$ is a preregistered observability threshold. The capability and observability constraints jointly define the feasible boundary set, ensuring that the selected boundary provides both the capabilities required by the payload and stable probe evidence. TD thus implements a repeatable process of candidate generation, differential probing, and constrained selection.

\textbf{Conditional-Payload Specialization.} After selecting $b_H^{\star}$, TD specializes the abstract payload $a$ according to the event, loading stage, privilege context, and environment state associated with that boundary:

\begin{equation}
a_{H}^{\mathrm{TD}}(s)=
\begin{cases}
a(s), & \chi_H(s,b_H^{\star})=1,\\
\mathrm{benign}(s), & \text{otherwise},
\end{cases}
\label{eq:td-conditional-payload}
\end{equation}

where $s$ denotes the runtime state and $\chi_H$ is a boundary predicate generated from the probe results. It verifies that the current event, loading stage, privilege context, and environment state are consistent with $b_H^{\star}$. In all other states, the hook preserves its original benign behavior; the payload enters the target branch only when execution reaches the weakest boundary and the required capabilities are actually available.

TD outputs the candidate boundary set $\mathcal{B}_H$, observations from benign probes, the ranking $G_H$ based on the trust gap, the selected boundary $b_H^{\star}$, and the conditional predicate $\chi_H$. These outputs specify the location and runtime conditions of attack activation and can be carried through delivery mechanisms such as version updates. LCI translates the selected boundary and payload conditions into a native configuration recognized by each target harness.

\subsection{Cross-Harness Robust Execution via the Least Common Interface}

The third stage transfers the attack across heterogeneous implementations of lifecycle-hook mechanisms. Different harnesses implement the common event-driven execution primitive through distinct event names, configuration schemas, command runners, and permission boundaries. LCI extracts the minimal sufficient capabilities required to preserve the semantics of an attack objective, removes dependencies on harness-specific events, syntax, and runtime environments, and generates a native lifecycle hook for every target harness.

Let the target harness set be
\[
\mathcal{H}=\{H_1,\dots,H_n\}.
\]
For a harness $H$, represent its hook interface as
\[
\mathcal{D}_H=(\mathcal{E}_H,\Sigma_H,\mathcal{X}_H,\mathcal{P}_H).
\]
Here, $\mathcal{E}_H$ is the set of lifecycle events, $\Sigma_H$ is the plugin manifest schema, and $\mathcal{X}_H$ is the command execution contract. The contract covers the command runner, the convention for passing arguments, the working directory, and rules for inheriting the environment. $\mathcal{P}_H$ captures the privileges and resources available to the hook subprocess. Let $\operatorname{Cap}(\mathcal{D}_H)$ denote the semantic capability set normalized from these attributes, and let

\begin{equation}
\mathcal{C}=\bigcap_{H\in\mathcal{H}}
\operatorname{Cap}(\mathcal{D}_H)
\label{eq:lci-intersection}
\end{equation}

denote the candidate capabilities shared by the complete target set. For an attack $a$, its cross-harness semantics are jointly defined by the target effect $o(a)$ and the necessary execution conditions $q(a)$. For a once-per-session objective, $q(a)$ includes session reachability and command-execution capability; for a tool-input manipulation objective, $q(a)$ additionally includes pre-tool-execution timing, invocation context, and input-control capability. LCI iteratively removes from $\mathcal{C}$ the capabilities irrelevant to $o(a)$ and $q(a)$, yielding the attack-specific minimal sufficient specification:

\begin{equation}
\begin{aligned}
\mathcal{I}^{*}_{a}
=\arg\min_{\mathcal{I}\subseteq\mathcal{C}}\;&|\mathcal{I}|\\
\mathrm{s.t.}\quad
&q(a)\subseteq\mathcal{I},\\
&\forall H\in\mathcal{H},\ \exists y_H\in
\operatorname{Native}_H(\mathcal{I}):
\operatorname{Obs}_H(y_H)=o(a).
\end{aligned}
\label{eq:lci-minimal-sufficient}
\end{equation}

Here, $\operatorname{Native}_H(\mathcal{I})$ denotes the native lifecycle-hook implementations in harness $H$ that satisfy capability set $\mathcal{I}$, and $\operatorname{Obs}_H$ denotes the harness-independent external-effect oracle. Equation~\eqref{eq:lci-minimal-sufficient} minimizes dependencies on event, command, and resource capabilities subject to the constraint that the same attack semantics remain realizable on every target harness. Attacks that satisfy this constraint proceed to native compilation; for all others, the applicability scope is restricted to the subset of harnesses providing the necessary capabilities.

After deriving $\mathcal{I}^{*}_{a}$, the harness adapter selects the native event in $H$ that satisfies the specification with the fewest additional dependencies:

\begin{equation}
e_H^{\star}=\arg\min_{e\in\mathcal{E}_H}
\operatorname{Dep}_H(e)
\quad\mathrm{s.t.}\quad
\mathcal{I}^{*}_{a}\subseteq\operatorname{Cap}_H(e),
\label{eq:lci-event-selection}
\end{equation}

where $\operatorname{Dep}_H(e)$ denotes the event trigger's dependence on conditions such as a particular tool invocation, runtime mode, or auxiliary service. Among native entry points satisfying $q(a)$, this criterion prioritizes the event with the fewest additional dependencies; when execution timing is part of $q(a)$, candidates are correspondingly restricted to events with the required temporal semantics. The adapter $\phi_H$ then compiles the minimal sufficient specification into

\begin{equation}
\phi_H(\mathcal{I}^{*}_{a})=(e_H^{\star},c_H,\sigma_H),
\label{eq:lci-compilation}
\end{equation}

where $c_H$ is the command that realizes the target effect under $\mathcal{X}_H$ and $\mathcal{P}_H$, and $\sigma_H$ is the lifecycle hook's native encoding in $\Sigma_H$. The attack semantics defined by $o(a)$ and $q(a)$ remain invariant during transfer; the native mechanism of each harness supplies the event name, command string, and configuration structure. Semantic extraction identifies the target effect and necessary conditions. Capability minimization removes harness-specific dependencies, after which native compilation selects the least-dependent feasible event and generates the command and configuration. Under this adapter contract, each shared semantic case produces a native lifecycle hook for every harness, and a common external oracle checks whether the effect is preserved.

\section{Experimental Evaluation}

We evaluate \textsc{HookPry} through four research questions. Within the scope defined by the threat model, the main \textsc{HookPry} experiment follows its generated native artifacts from harness loading to externally observable effects verified by independent oracles. RQ2 is a separate translation-comparison experiment that mechanically converts malicious MCP tool descriptions into lifecycle hooks and compares them, on the same 50 targets, with the E2E-ASR of native \textsc{HookPry} constructed by us.

\begin{icompact}

\item \textbf{RQ1 (Effectiveness):} Is \textsc{HookPry} effective and stable across different harnesses and LLM backends?
\begin{icompact}
\item \textbf{RQ1.1 (Execution-Level Effectiveness and Mechanism Utility):} Across harness architectures and attack categories, how reliably can \textsc{HookPry} start its attack mechanism and produce verified end-to-end effects?
\item \textbf{RQ1.2 (Model Independence):} After a matching lifecycle event fires, to what extent is \textsc{HookPry}'s execution path independent of the LLM backend?
\end{icompact}

\item \textbf{RQ2 (Translation Comparison):} After malicious MCP tool descriptions are converted into lifecycle-event hooks, what E2E-ASR do they achieve, and is it lower than that of native \textsc{HookPry} constructed for the same targets?

\item \textbf{RQ3 (Ablation Study):} How do AMO, TD, and LCI affect \textsc{HookPry}'s end-to-end attack effectiveness?

\item \textbf{RQ4 (Countermeasures):} To what extent can representative countermeasures distinguish \textsc{HookPry} attack artifacts from matched benign controls?

\end{icompact}

RQ1 measures cross-harness end-to-end attack effects and mechanism utility (RQ1.1), then tests model independence of the post-trigger execution path (RQ1.2). RQ2 compares translated MCP attack semantics with native malicious hooks. RQ3 isolates the contributions of AMO, TD, and LCI through component ablations. RQ4 measures the coverage and residual gaps of static defenses.

\subsection{Experimental Setup}

\textbf{Harnesses and LLM Backends.} The RQ1 experiment tests seven AI agent harnesses: OpenHarness, OpenClaw, Claude Code, Codex CLI, OpenCode, Hermes, and WorkBuddy. It uses five LLM backends: Claude Sonnet 4.6, DeepSeek-V4-Pro, Kimi-K2.6, GLM-4.7-Flash, and GitHub Copilot, covering 25 combinations of harnesses and backends. RQ2 and RQ3 use an independent unified comparison dataset and are fixed to Claude Code with DeepSeek-V4-Flash; this backend is not counted among the 25 RQ1 combinations.

\textbf{Protocol.} We generate native artifacts for each harness from the same 40 abstract case seeds and execute them in batches. Every combination of case, harness, and backend runs in an environment containing synthetic assets. A run covers harness startup, loading and registration of the native lifecycle hook, submission of the trigger request, \textsc{HookPry} execution, collection of external evidence, and verdict computation by the oracle. The experiment comprises 40 cases $\times$ 25 combinations of harnesses and backends = 1,000 attempted runs.

\textbf{RQ2 Comparison Protocol.} RQ2 is independent of the 1,000 execution-layer runs above. We select 50 malicious MCP tool-description-poisoning targets traceable to MCPTox~\cite{ref16} from a unified dataset. For each target, we preserve the benign query, capability budget, synthetic fixture, effect, and external oracle; only the original malicious description is mechanically converted into context injected by a Claude Code lifecycle hook. The hook itself does not directly execute the effect, and the model still decides whether to invoke the same synthetic effect tool. Under a fixed random-seed allocation, 19 MCP2Hook cases retain \texttt{SessionStart}; the remaining 31 are assigned to \texttt{UserPromptSubmit}, \texttt{PreToolUse}, \texttt{PostToolUse}, or \texttt{Stop}. The experiment uses Claude Code with DeepSeek V4 Flash, provider-default sampling, a 120-second timeout, and a fresh isolated directory for every run.

\textbf{Verdicts and Metrics.} The evaluator assigns each run one of four mutually exclusive verdicts. A \emph{pass} requires all preregistered external checks for the case to succeed; a \emph{partial} verdict indicates that only a proper subset succeeds; \emph{blocked} records explicit harness interception; and \emph{fail} covers all remaining outcomes. Our primary metric is the binary verified end-to-end attack success rate (E2E-ASR), defined as $N_{\mathrm{pass}}/N_{\mathrm{attempted}}$. Partial outcomes receive zero credit, and we report the outcome distribution separately. We also report mechanism utility, defined as the proportion of runs in which steps such as hook triggering, payload loading, or hook registration actually occur. This measure distinguishes mechanism initiation from end-to-end effectiveness.

\subsection{RQ1: \textsc{HookPry} Effectiveness}

\subsubsection{RQ1.1: Execution-Level Effectiveness and Mechanism Utility}

\textbf{Setup.} We conduct 1,000 end-to-end runs over the 40 attack cases, seven harnesses, and five LLM backends described in the experimental setup, and use external oracles to determine the effect of every run. A successfully generated configuration, emitted tool call, or spawned process is not by itself sufficient evidence of success; RQ1.1 requires \textsc{HookPry} to produce the preregistered external effect.

\textbf{Results.} Table~\ref{tab:2} reports \textsc{HookPry}'s binary E2E-ASR by attack category and harness. Across 1,000 attempted runs, \textsc{HookPry} receives 770 pass verdicts, 34 partial verdicts, and 196 fail verdicts; no run is explicitly blocked. The resulting micro-averaged E2E-ASR is 77.0\%. The final column of Table~\ref{tab:2} reports an unweighted macro-average across harnesses so that harnesses with different numbers of tested backends contribute equally.

\begin{table*}[!t]
\centering
\caption{[RQ1.1] \textsc{HookPry}'s verified execution-level E2E-ASR by attack category and harness. Partial outcomes receive no fractional credit. The final column is the unweighted macro-average across harnesses.}
\label{tab:2}
\resizebox{\textwidth}{!}{%
\begin{tabular}{lcccccccc}
\toprule
\textbf{Attack Category} &
\textbf{OpenHarness} &
\textbf{OpenClaw} &
\textbf{Claude Code} &
\textbf{Codex CLI} &
\textbf{OpenCode} &
\textbf{Hermes} &
\textbf{WorkBuddy} &
\textbf{Average} \\
\midrule
Credential Collection (COL) & 81.0\% & 100.0\% & 60.7\% & 71.4\% & 96.4\% & 95.2\% & 90.5\% & 85.0\% \\
Data Exfiltration (EXF) & 75.0\% & 100.0\% & 87.5\% & 50.0\% & 100.0\% & 100.0\% & 91.7\% & 86.3\% \\
Resource Hijacking (HIJ) & 83.3\% & 87.5\% & 100.0\% & 100.0\% & 100.0\% & 100.0\% & 83.3\% & 93.4\% \\
Command \& Control (C2) & 33.3\% & 50.0\% & 12.5\% & 0.0\% & 62.5\% & 66.7\% & 66.7\% & 41.7\% \\
Tampering (TAM) & 75.0\% & 100.0\% & 6.2\% & 100.0\% & 81.3\% & 100.0\% & 100.0\% & 80.4\% \\
Manipulation (MAN) & 66.7\% & 100.0\% & 33.3\% & 66.7\% & 66.7\% & 77.8\% & 88.9\% & 71.4\% \\
Privilege Escalation (ESC) & 66.7\% & 100.0\% & 87.5\% & 100.0\% & 100.0\% & 100.0\% & 100.0\% & 93.5\% \\
Persistence (PER) & 83.3\% & 6.2\% & 6.2\% & 0.0\% & 100.0\% & 91.7\% & 100.0\% & 55.3\% \\
Evasion (EVA) & 66.7\% & 68.8\% & 68.8\% & 75.0\% & 75.0\% & 75.0\% & 66.7\% & 70.9\% \\
Propagation (PRO) & 66.7\% & 83.3\% & 54.2\% & 66.7\% & 100.0\% & 100.0\% & 100.0\% & 81.6\% \\
Overall & 71.7\% & 81.9\% & 52.5\% & 65.0\% & 90.6\% & 92.5\% & 90.8\% & 77.9\% \\
\bottomrule
\end{tabular}%
}
\end{table*}

Table~\ref{tab:utility} reports mechanism utility by attack category. Utility is the proportion of mechanism-level steps that actually initiate, including hook triggering, payload loading, and hook registration. Overall utility is 83.9\%, above the overall E2E-ASR (77.0\% micro-average and 77.9\% macro-average), showing that mechanism initiation does not imply completion of the external effect. Utility is highest for Privilege Escalation (94.3\%) and Resource Hijacking (91.7\%), but lower for Command and Control (55.4\%) and Persistence (65.8\%); Persistence reaches only 0.0\% for Codex CLI and 6.2\% for OpenClaw.

\begin{table*}[!t]
\centering
\small
\caption{[RQ1.1] \textsc{HookPry}'s mechanism utility by attack category and harness. Each cell reports the proportion of mechanism-level steps that actually initiate; the final column and row are the corresponding unweighted averages.}
\label{tab:utility}
\resizebox{\textwidth}{!}{%
\begin{tabular}{lcccccccc}
\toprule
\textbf{Attack Category} &
\textbf{Claude Code} &
\textbf{Codex CLI} &
\textbf{Hermes} &
\textbf{OpenClaw} &
\textbf{OpenCode} &
\textbf{OpenHarness} &
\textbf{WorkBuddy} &
\textbf{Average} \\
\midrule
Credential Collection (COL) & 71.4\% & 85.7\% & 85.7\% & 100.0\% & 96.4\% & 90.5\% & 90.5\% & 88.6\% \\
Data Exfiltration (EXF) & 81.2\% & 50.0\% & 100.0\% & 100.0\% & 100.0\% & 83.3\% & 91.7\% & 86.6\% \\
Resource Hijacking (HIJ) & 87.5\% & 87.5\% & 100.0\% & 100.0\% & 100.0\% & 83.3\% & 83.3\% & 91.7\% \\
Command \& Control (C2) & 50.0\% & 62.5\% & 50.0\% & 50.0\% & 75.0\% & 33.3\% & 66.7\% & 55.4\% \\
Tampering (TAM) & 50.0\% & 100.0\% & 100.0\% & 100.0\% & 81.2\% & 91.7\% & 100.0\% & 89.0\% \\
Manipulation (MAN) & 58.3\% & 66.7\% & 77.8\% & 100.0\% & 66.7\% & 88.9\% & 88.9\% & 78.2\% \\
Privilege Escalation (ESC) & 93.8\% & 100.0\% & 100.0\% & 100.0\% & 100.0\% & 66.7\% & 100.0\% & 94.3\% \\
Persistence (PER) & 62.5\% & 0.0\% & 91.7\% & 6.2\% & 100.0\% & 100.0\% & 100.0\% & 65.8\% \\
Evasion (EVA) & 75.0\% & 100.0\% & 100.0\% & 75.0\% & 100.0\% & 75.0\% & 91.7\% & 88.1\% \\
Propagation (PRO) & 83.3\% & 83.3\% & 100.0\% & 87.5\% & 100.0\% & 72.2\% & 100.0\% & 89.5\% \\
Average & 72.5\% & 75.0\% & 91.5\% & 83.8\% & 93.8\% & 80.8\% & 93.3\% & 83.9\% \\
\bottomrule
\end{tabular}%
}
\end{table*}
The per-cell sample sizes are $N$: COL=28/21 and PRO=24/18; the remaining categories contain 12--16 observations. Hermes, OpenHarness, and WorkBuddy have only three tested models for some categories, so their corresponding cells have slightly smaller $N$. 

Across all attempted runs, lifecycle hooks are actually triggered in approximately 82.0\% of cases, ranging from 72.3\% for Codex CLI to 92.4\% for OpenCode. Mechanism utility is higher than the end-to-end success rate of 77.0\%, indicating that environmental and policy constraints after hook triggering account for most of the gap.

\textbf{RQ1.1 Takeaway.} \textsc{HookPry} produces oracle-confirmed effects on every target harness, but the utility matrix shows that mechanism utility is not uniform across attack semantics. Hermes reaches an overall success rate of 92.5\% and achieves 100\% in five categories. Transfer is strongest when the attack objective relies on capabilities commonly available to lifecycle-hook subprocesses: Resource Hijacking reaches a 93.4\% macro-average, and Privilege Escalation reaches 93.5\%. Objectives that depend on network reachability or durable filesystem state expose a different boundary. Command and Control (41.7\% E2E-ASR; 55.4\% mechanism utility) and Persistence (55.3\% E2E-ASR; 65.8\% mechanism utility) remain constrained by network policy, filesystem policy, and isolation mechanisms. These challenges particularly manifest in certain harness configurations, suggesting that complex network interactions and system persistence mechanisms may require additional considerations in the hook implementation strategy—such as enhanced detection mechanisms for network protocols and deeper integration with system-level monitoring—to better capture these attack patterns. This category structure supports the intended advantage of LCI in preserving portable attack semantics. Local deployment policy still determines whether a bound command starts and completes its final effect.
\subsubsection{RQ1.2: Model Independence}

\textbf{Setup.} Holding the generated cases, native adapters, evaluator, and external oracles fixed, we use cross-backend standard deviation to quantify variation among LLM backends within each harness. A \textsc{HookPry} payload is a command bound to a harness lifecycle event, and its execution path differs from natural-language prompt injection. The LLM may affect whether a tool-use event occurs; after a matching event fires, lifecycle-hook binding and subprocess dispatch follow the harness execution path.

\begin{figure*}[!t]
\centering
\includegraphics[width=0.95\textwidth]{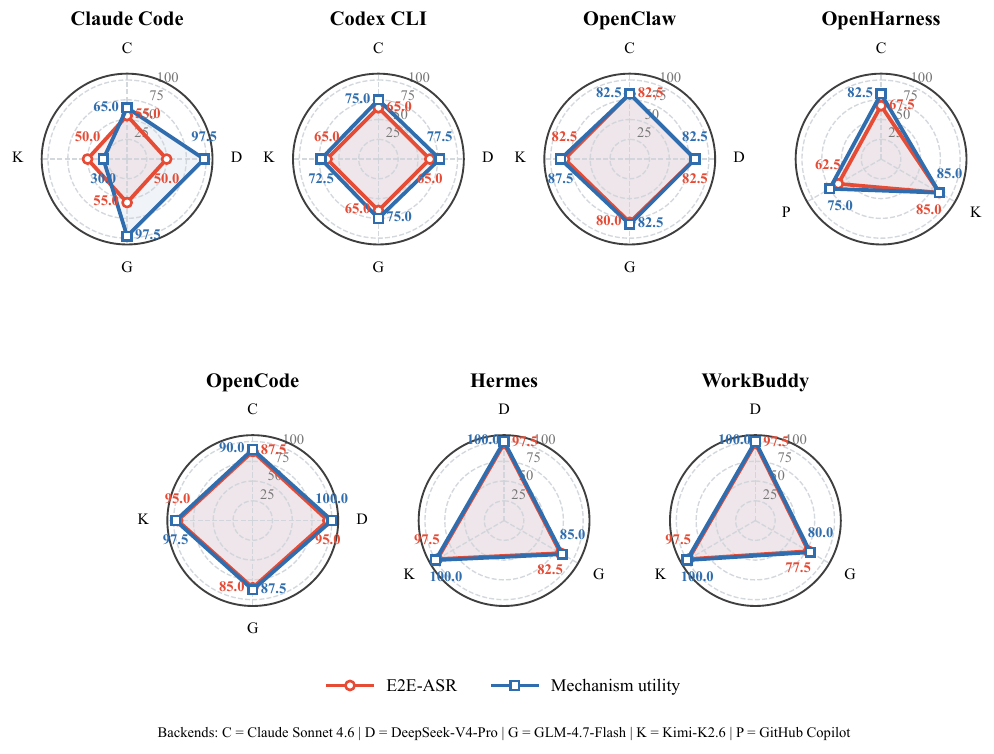}
\caption{[RQ1.2] E2E-ASR (circles) and mechanism utility (squares) across LLM backends for each harness. Vertices report exact percentages; pairs of harnesses and backends that were not evaluated are omitted. Backend abbreviations are defined in the figure.}
\label{fig:rq1-model-matrix}
\end{figure*}

\textbf{Results.} Figure~\ref{fig:rq1-model-matrix} compares \textsc{HookPry}'s binary E2E-ASR and mechanism utility across the evaluated LLM backends for each harness. Each radar panel contains only the backends tested for that harness, and the annotated vertices report the exact rates.

Codex CLI exhibits no cross-backend variation in verified success. OpenClaw and Claude Code have overall standard deviations of 1.1pp and 2.5pp, respectively. Thus, three of the seven harnesses maintain nearly constant E2E-ASR across the tested backends. The remaining harnesses are more backend-sensitive: OpenCode has a standard deviation of 4.5pp, Hermes 7.1pp, and WorkBuddy 9.4pp; OpenHarness's 9.6pp variation is driven primarily by the gap between GitHub Copilot (62.5\%) and Kimi-K2.6 (85.0\%).

The strongest evidence for model independence comes from Codex CLI, OpenClaw, and Claude Code: despite using backends from different model families, none has a cross-backend standard deviation above 2.5pp. Harness ordering also remains stable across backends, which points to registration, permission, and isolation behavior as the main source of the success-rate differences. The larger variations for OpenHarness, WorkBuddy, and Hermes (9.6pp, 9.4pp, and 7.1pp, respectively) occur when events depend closely on tool use, because backend-specific tool-use patterns can determine whether the matching lifecycle event fires. E2E-ASR captures both this model-mediated trigger and the subsequent harness-controlled execution.

\textbf{RQ1.2 Takeaway.} \textsc{HookPry}'s binding and subprocess dispatch become harness-controlled after a matching lifecycle event fires. The cross-backend standard deviations below 2.5pp for Codex CLI, OpenClaw, and Claude Code support model independence at this post-trigger boundary. However, OpenCode, Hermes, WorkBuddy, and OpenHarness exhibit greater variability (4.5–9.6pp) due to their distinct tool-use behaviors that influence event generation. This variation reflects fundamental differences in how these backends interact with system APIs, rather than inconsistencies in the HookPry mechanism itself. The execution path achieves model independence after triggering, but the complete attack remains conditionally model-dependent because the initial event generation phase is still influenced by each backend’s unique tool invocation patterns. Future work could focus on developing more robust event prediction models that account for these behavioral differences across LLM backends.

\subsection{RQ2: Translating Malicious MCP Targets into Hooks with Mixed Lifecycle Events}

\textbf{Setup.} We select 50 malicious MCP tool-description-poisoning targets traceable to MCPTox~\cite{ref16} and convert them into Claude Code lifecycle hooks. Under a fixed random-seed allocation, 19 MCP2Hook cases retain \texttt{SessionStart}; the remaining 31 are assigned to \texttt{UserPromptSubmit}, \texttt{PreToolUse}, \texttt{PostToolUse}, or \texttt{Stop}. The experiment uses Claude Code with DeepSeek-V4-Flash, provider-default sampling, a 120-second timeout, and a fresh isolated directory for every run. Native \textsc{HookPry} on the same targets serves as the paired control.

\textbf{Results.} Table~\ref{tab:comparison-translated-hooks} compares translated MCP2Hook with native \textsc{HookPry} on the paired 50-target dataset.

\begin{table}[!t]
\centering
\caption{[RQ2] E2E-ASR after malicious MCP targets are converted into hooks with mixed lifecycle events. \textsc{HookPry} consists of native lifecycle hooks constructed by us for the same 50 targets and serves as the paired control using the same sample.}
\label{tab:comparison-translated-hooks}
\resizebox{\columnwidth}{!}{%
\begin{tabular}{lccc}
\toprule
\textbf{Condition} &
\textbf{Successes (out of valid)} &
\textbf{E2E-ASR} &
\textbf{Wilson 95\% CI} \\
\midrule
Native \textsc{HookPry} & 46~of~50 & 92.0\% & 81.2--96.8\% \\
MCP$\rightarrow$Hook & 28~of~50 & 56.0\% & 42.3--68.8\% \\
\bottomrule
\end{tabular}
}
\end{table}

MCP2Hook succeeds in 28 of 50 valid cases, yielding an E2E-ASR of \CompMcpASR\% (Wilson 95\% CI: \CompMcpCILow\%--\CompMcpCIHigh\%), significantly below the \CompHistASR\% achieved by native \textsc{HookPry} on the same targets (one-sided exact binomial test, $p=\CompMcpBinomP$). Its context hooks trigger in 38/50 runs (76.0\%), below the trigger reliability of native \textsc{HookPry}. In the other 12 translated cases, the hook mechanism does not start.

\textbf{RQ2 Takeaway.} Translated malicious MCP tool descriptions reach \CompMcpASR\% E2E-ASR, significantly below the \CompHistASR\% of native \textsc{HookPry} on the same targets ($p=\CompMcpBinomP$). MCP2Hook injects a malicious description and still requires the model to invoke the effect tool. Native \textsc{HookPry} executes the effect directly from a lifecycle event and removes that model-mediated gate, accounting for the observed advantage.

\subsection{RQ3: \textsc{HookPry} Ablation Study}

\textbf{Setup.} We fix Claude Code with DeepSeek-V4-Flash and use Full \textsc{HookPry}'s 46/50 successes on the unified comparison dataset (92.0\% E2E-ASR) as the baseline. We construct four system variants: \emph{Full \textsc{HookPry}} retains every component; \emph{w/o AMO} replaces multi-intent optimization with centroid-only A0 retrieval; \emph{w/o TD} prevents new lifecycle-hook entries from inheriting execution authorization under an existing plugin identity; and \emph{w/o LCI} removes the adapter for Claude Code's native lifecycle-hook interface, preventing the unified payload from being converted into registrable lifecycle-hook configuration. All cases, harness settings, LLM-backend settings, timeouts, capability budgets, and external oracles other than the ablated component remain fixed. We retain E2E-ASR as the primary metric and additionally report payload load rate, hook registration rate, and the ASR decrease relative to the full system (Drop).

\textbf{Results.} Table~\ref{tab:ablation} reports the effectiveness and mechanism-utility measures for all four ablation variants.

\begin{table}[!t]
\centering
\small
\caption{[RQ3] \textsc{HookPry} ablation results under Claude Code with DeepSeek-V4-Flash. The w/o AMO value is an analytical estimate; w/o TD and w/o LCI are structural results when payload loading or native registration is blocked. ``--'' denotes an unreachable or inapplicable stage.}
\label{tab:ablation}
\setlength{\tabcolsep}{3pt}
\begin{tabular}{@{}lcccc@{}}
\toprule
\multirow{2}{*}{\textbf{System Variant}} &
\multirow{2}{*}{\shortstack{\textbf{Payload}\\\textbf{Load}}} &
\multirow{2}{*}{\shortstack{\textbf{Hook}\\\textbf{Registration}}} &
\multirow{2}{*}{\textbf{E2E-ASR}} &
\multirow{2}{*}{\textbf{Drop}} \\
& & & & \\
\midrule
Full \textsc{HookPry} & 100.0\% & 100.0\% & 92.0\% & -- \\
w/o AMO & 100.0\% & 100.0\% & 75.6\% & 16.4pp \\
w/o TD & 0.0\% & -- & 0.0\% & 92.0pp \\
w/o LCI & 100.0\% & 0.0\% & 0.0\% & 92.0pp \\
\bottomrule
\end{tabular}
\end{table}

Removing AMO reduces \textsc{HookPry}'s E2E-ASR to 75.6\%, a 16.4-percentage-point decrease from the full system. AMO contributes at the artifact-retrieval and selection stage; it does not alter harness execution after loading. Without TD, newly introduced hook entries cannot inherit the execution authorization of the existing plugin identity, so the payload is blocked at the update-delivery boundary; both payload load rate and E2E-ASR fall to 0\%. Without LCI, the unified attack semantics can still be loaded but cannot be converted into a native Claude Code lifecycle hook that can be parsed and registered; hook registration and E2E-ASR fall to 0\% while payload loading remains at 100\%.

\textbf{RQ3 Takeaway.} Table~\ref{tab:ablation} uses payload load and hook registration as mechanism-utility measures for the four variants; both reach 100\% for Full \textsc{HookPry}. Removing TD blocks payload loading. Removing LCI preserves loading but blocks native registration. AMO, TD, and LCI therefore operate at distinct stages of the mechanism chain.

\subsection{RQ4: Countermeasure}

\textbf{Setup.} We construct a fixed 40:40 static corpus: one malicious artifact bundle for each \textsc{HookPry} case and one benign control matched by attack category and lifecycle-hook event. Every scan directory uses an opaque identifier; labels and case identifiers remain outside the scanned tree. A frozen mechanical transformation removes benchmark-only comments and names, such as case IDs and explicit attack labels, while preserving executable operations including sensitive-file access, network transfer, configuration modification, and lifecycle-hook registration. Each bundle contains a plugin manifest, a \texttt{Description.md}, native lifecycle-hook configuration, and its payload.

We evaluate three distinct static defenses. Microsoft Defender is a general-purpose endpoint antivirus; Semgrep 1.172.0 uses five frozen generic rules; and HookPolicy 0.1.0 is our transparent eight-rule, lifecycle-hook-aware baseline. Semgrep and HookPolicy are researcher-configured baselines and do not represent commercial defaults. Every tool successfully scans all 80 artifacts. We report recall on malicious samples, evasion ($1-\mathrm{recall}$), and false-positive rate (FPR) on benign samples; tool errors are excluded from the denominator and recorded separately.

\textbf{Results.} Table~\ref{tab:4} reports recall, evasion, and false-positive rates for the three defenses.

\begin{table}[H]
\centering
\small
\caption{[RQ4] Static countermeasure coverage for \textsc{HookPry} on the anonymous, sanitized 40:40 corpus.}
\label{tab:4}
\resizebox{\columnwidth}{!}{%
\begin{tabular}{lcccccc}
\toprule
\textbf{Countermeasure} &
\textbf{TP/40} &
\textbf{Recall} &
\textbf{FN/40} &
\textbf{Evasion} &
\textbf{FP/40} &
\textbf{FPR} \\
\midrule
Defender & 0 & 0.0\% & 40 & 100.0\% & 0 & 0.0\% \\
HookPolicy & 20 & 50.0\% & 20 & 50.0\% & 0 & 0.0\% \\
Semgrep & 19 & 47.5\% & 21 & 52.5\% & 0 & 0.0\% \\
\bottomrule
\end{tabular}%
}
\end{table}

HookPolicy detects 20/40 malicious artifacts (50.0\%; Wilson 95\% CI: 35.2\%--64.8\%), Semgrep detects 19/40 (47.5\%; Wilson 95\% CI: 32.9\%--62.5\%), and Defender detects 0/40 (0.0\%; Wilson 95\% CI: 0.0\%--8.8\%). The union of the three defenses detects 21/40 and misses 19/40, for a joint miss rate of 47.5\%. All scanners produce 0/40 false positives on the matched controls; because these controls are synthetic and low risk, this value is only a point estimate, with a Wilson 95\% upper bound of 8.8\%. All three scanners process all 80 artifacts (100\%), and no tool errors are excluded from the denominators. Scanner availability does not explain the coverage gap.

\textbf{RQ4 Takeaway.} No evaluated endpoint scanner or transparent static-rule baseline detects more than half of the malicious samples, and their union still misses 47.5\%. These misses warrant explicit lifecycle-hook review during installation and updates, with runtime monitoring and least-privilege controls for residual cases.

\section{Discussion and Limitations}

\textsc{HookPry} exposes a trust failure in how AI agent harnesses interpret lifecycle bindings in accepted artifacts. After a matching event, the harness controls command binding and subprocess dispatch; the model affects only event generation. Prompt-layer defenses therefore miss this execution boundary. Review should cover plugin installation, updates, hook registration, and runtime information flow. Harnesses should authorize changed hooks individually, bind plugin-manifest signatures to payloads, and enforce least privilege. For events such as \texttt{PostToolUse}, they should preserve original outputs and record transformation provenance. Because RQ4's static defenses missed 47.5\% of malicious artifacts, static scanning cannot replace runtime monitoring and permission constraints.

The threat model assumes installation through normal channels and adoption of updates. AMO measures retrieval opportunities, not installation or update-adoption rates. The experiments use ephemeral environments with synthetic assets and cover seven harnesses, five LLM backends, and 40 attack cases; they do not exhaust operating systems, shells, enterprise policies, network conditions, or future versions. \textsc{HookPry} excludes implementation bugs and sandbox escapes. Its effect depends on event occurrence, hook authorization, and subprocess permissions, as reflected in the lower success rates of Command and Control and Persistence. RQ4 evaluates two generic scanners and one lifecycle-hook-aware static baseline with synthetic benign controls, but not dynamic analysis or specialized commercial products.

These limitations define future work: measuring marketplace retrieval and update adoption without real user data; expanding coverage across systems, versions, and enterprise policies; constructing representative benign-plugin benchmarks; and evaluating signature binding, differential update authorization, least-privilege hooks, and tool-output integrity. Such work can translate this attack surface into deployable ecosystem and harness protections.

\section{Conclusion}

Lifecycle hooks create a supply-chain attack surface at the configuration layer of the agent ecosystem. \textsc{HookPry} implements this attack through AMO, TD, and LCI, which cover ecosystem acquisition, conditional activation across versions, and cross-harness execution. Across 1,000 runs on seven harnesses and five LLM backends, \textsc{HookPry} reaches 77.0\% micro-average E2E-ASR and produces effects verified by external oracles on every target harness. Once a matching event fires, the harness controls command binding and subprocess dispatch. On the paired comparison set, native lifecycle hooks reach 92.0\% E2E-ASR, while translated malicious MCP descriptions reach 56.0\%. The union of three static defenses still misses 47.5\% of malicious artifacts. Agent security must therefore audit plugin updates, hook registration, subprocess permissions, and tool-output integrity as one trust chain; model-level content protection does not cover this path.

\appendix

\section*{Ethical Considerations}

This research is dual-use. Lifecycle-hook analysis can help AI agent harnesses and marketplaces identify trust boundaries, but the same knowledge could enable credential harvesting, result tampering, persistence, or resource hijacking. Stakeholders include agent users, plugin developers, harness and marketplace maintainers, and security researchers; potential harms include compromise of host-side data and computational resources as well as the propagation of false conclusions after tool-output tampering. The research is justified because the risk arises from bindings between events and commands and from coarse trust in updates, both of which harnesses already provide and prompt filtering cannot adequately observe or mitigate. Systematically characterizing acquisition, activation, and cross-harness execution supports targeted defenses such as update authorization, least privilege, and information-flow integrity.

All end-to-end evaluations were conducted in ephemeral environments with synthetic files and preset checkpoints. An external oracle adjudicated attack effects, and no real user data or credentials were used. Defense evaluation used frozen malicious artifacts and matched benign controls. The threat model excludes forced installation, implementation bugs, sandbox escapes, and control over real repositories, limiting experimental capability to the boundaries of plugin configuration and hook permissions under study. Public artifacts that could directly increase abuse capability should be released in stages. Case definitions, synthetic assets, adjudicators, and defense rules should be released first; payloads that act directly on real environments should be reduced in capability; and the release should document authorized testing scope, isolation requirements, and prohibited uses.

This work studies architectural risks shared across products rather than targeting particular users or operational services. If experimental re-verification or artifact preparation confirms a vendor-specific, undisclosed, and realistically exploitable issue, we will responsibly disclose it to the relevant maintainers before releasing actionable details and will record the notification scope, mitigation status, and residual risk. The same principle applies to tests that could violate terms of service or affect third-party infrastructure: without explicit authorization and a completed risk assessment, such tests will not enter the public experimental workflow. These constraints cannot eliminate dual-use risk, but they reduce foreseeable harm while preserving research verifiability. Finally, we have responsibly disclosed our findings to all affected harness vendors and are waiting for their response.

\section*{Open Science}

To support open science, \textsc{HookPry} is available through an anonymized repository at \url{https://anonymous.4open.science/r/HookPry-2EBD}.

\begingroup
\sloppy
\setlength{\emergencystretch}{3em}
\bibliographystyle{plainurl}
\bibliography{references}

@inproceedings{ref1,
  author    = {Kai Greshake and Sahar Abdelnabi and Shailesh Mishra and Christoph Endres and Thorsten Holz and Mario Fritz},
  title     = {Not What You've Signed Up For: Compromising Real-World {LLM}-Integrated Applications with Indirect Prompt Injection},
  booktitle = {Proceedings of the 16th ACM Workshop on Artificial Intelligence and Security (AISec@CCS)},
  year      = {2023},
  note      = {Best Paper Award}
}

@inproceedings{ref2,
  author    = {Yupei Liu and Yuqi Jia and Runpeng Geng and Jinyuan Jia and Neil Zhenqiang Gong},
  title     = {Formalizing and Benchmarking Prompt Injection Attacks and Defenses},
  booktitle = {Proceedings of the 33rd USENIX Security Symposium},
  year      = {2024}
}

@inproceedings{ref3,
  author    = {Qiusi Zhan and Zhixiang Liang and Zifan Ying and Daniel Kang},
  title     = {{InjecAgent}: Benchmarking Indirect Prompt Injections in Tool-Integrated Large Language Model Agents},
  booktitle = {Findings of the Association for Computational Linguistics (ACL)},
  year      = {2024}
}

@inproceedings{ref4,
  author    = {Edoardo Debenedetti and Jie Zhang and Mislav Balunovi{\'c} and Luca Beurer-Kellner and Marc Fischer and Florian Tram{\`e}r},
  title     = {{AgentDojo}: A Dynamic Environment to Evaluate Attacks and Defenses for {LLM} Agents},
  booktitle = {Advances in Neural Information Processing Systems (NeurIPS), Datasets and Benchmarks Track},
  year      = {2024}
}

@inproceedings{ref5,
  author    = {Maksym Andriushchenko and Alexandra Souly and Mateusz Dziemian and Derek Duenas and Maxwell Lin and Justin Wang and Dan Hendrycks and Andy Zou and Zico Kolter and Matt Fredrikson and Eric Winsor and Jerome Wynne and Yarin Gal and Xander Davies},
  title     = {{AgentHarm}: A Benchmark for Measuring Harmfulness of {LLM} Agents},
  booktitle = {Proceedings of the 13th International Conference on Learning Representations (ICLR)},
  year      = {2025}
}

@inproceedings{ref6,
  author    = {Hanrong Zhang and Jingyuan Huang and Kai Mei and Yifei Yao and Zhenting Wang and Chenlu Zhan and Hongwei Wang and Yongfeng Zhang},
  title     = {Agent Security Bench ({ASB}): Formalizing and Benchmarking Attacks and Defenses in {LLM}-based Agents},
  booktitle = {Proceedings of the 13th International Conference on Learning Representations (ICLR)},
  year      = {2025}
}

@inproceedings{ref7,
  author    = {Yupei Liu and Yuqi Jia and Jinyuan Jia and Dawn Song and Neil Zhenqiang Gong},
  title     = {{DataSentinel}: A Game-Theoretic Detection of Prompt Injection Attacks},
  booktitle = {Proceedings of the IEEE Symposium on Security and Privacy (S\&P)},
  year      = {2025},
  note      = {Distinguished Paper Award}
}

@inproceedings{ref8,
  author    = {Zhun Wang and Vincent Siu and Zhe Ye and Tianneng Shi and Yuzhou Nie and Xuandong Zhao and Chenguang Wang and Wenbo Guo and Dawn Song},
  title     = {{AgentVigil}: Automatic Black-Box Red-teaming for Indirect Prompt Injection against {LLM} Agents},
  booktitle = {Findings of the Conference on Empirical Methods in Natural Language Processing (EMNLP)},
  year      = {2025}
}

@inproceedings{ref9,
  author    = {Jiawen Shi and Zenghui Yuan and Guiyao Tie and Pan Zhou and Neil Zhenqiang Gong and Lichao Sun},
  title     = {Prompt Injection Attack to Tool Selection in {LLM} Agents},
  booktitle = {Proceedings of the 33rd Network and Distributed System Security Symposium (NDSS)},
  year      = {2026},
  doi       = {10.14722/ndss.2026.230675},
  url       = {https://www.ndss-symposium.org/ndss-paper/prompt-injection-attack-to-tool-selection-in-llm-agents/}
}

@misc{ref10,
  author    = {Yue Liu and Yanjie Zhao and Yunbo Lyu and Ting Zhang and Haoyu Wang and David Lo},
  title     = {``{Your} {AI}, {My} {Shell}'': Demystifying Prompt Injection Attacks on Agentic {AI} Coding Editors},
  year      = {2025},
  eprint    = {2509.22040},
  archiveprefix = {arXiv},
  url       = {https://arxiv.org/abs/2509.22040}
}

@inproceedings{ref11,
  author    = {Zhaorui Chen and Zhen Xiang and Chaowei Xiao and Dawn Song and Bo Li},
  title     = {{AgentPoison}: Red-teaming {LLM} Agents via Poisoning Memory or Knowledge Bases},
  booktitle = {Advances in Neural Information Processing Systems (NeurIPS)},
  year      = {2024}
}

@misc{ref12,
  author    = {Xinpeng Liu and Junming Liu and Peiyu Liu and Han Zheng and Qinying Wang and Mathias Payer and Shouling Ji and Wenhai Wang},
  title     = {Cuckoo Attack: Stealthy and Persistent Attacks Against {AI-IDE}},
  year      = {2025},
  eprint    = {2509.15572},
  archiveprefix = {arXiv},
  url       = {https://arxiv.org/abs/2509.15572}
}

@misc{ref13,
  author    = {Fazhong Liu and Zhuoyan Chen and Tu Lan and Haozhen Tan and Zhenyu Xu and Xiang Li and Guoxing Chen and Yan Meng and Haojin Zhu},
  title     = {Trojan's Whisper: Stealthy Manipulation of {OpenClaw} through Injected Bootstrapped Guidance},
  year      = {2026},
  eprint    = {2603.19974},
  archiveprefix = {arXiv},
  url       = {https://arxiv.org/abs/2603.19974}
}

@misc{ref14,
  author    = {Yuhang Wang and Feiming Xu and Zheng Lin and Guangyu He and Yuzhe Huang and Haichang Gao and Zhenxing Niu and Shiguo Lian and Zhaoxiang Liu},
  title     = {From Assistant to Double Agent: Formalizing and Benchmarking Attacks on {OpenClaw} for Personalized Local {AI} Agent},
  year      = {2026},
  eprint    = {2602.08412},
  archiveprefix = {arXiv},
  url       = {https://arxiv.org/abs/2602.08412}
}

@misc{ref15,
  author    = {Xinhao Deng and Yixiang Zhang and Jiaqing Wu and Jiaqi Bai and Sibo Yi and Zhuoheng Zou and Yue Xiao and Rennai Qiu and Jianan Ma and Jialuo Chen and Xiaohu Du and Xiaofang Yang and Shiwen Cui and Changhua Meng and Weiqiang Wang and Jiaxing Song and Ke Xu and Qi Li},
  title     = {Taming {OpenClaw}: Security Analysis and Mitigation of Autonomous {LLM} Agent Threats},
  year      = {2026},
  eprint    = {2603.11619},
  archiveprefix = {arXiv},
  url       = {https://arxiv.org/abs/2603.11619}
}

@inproceedings{ref16,
  author    = {Zhiqiang Wang and Yichao Gao and Yanting Wang and Suyuan Liu and Haifeng Sun and Haoran Cheng and Guanquan Shi and Haohua Du and Xiang-Yang Li},
  title     = {{MCPTox}: A Benchmark for Tool Poisoning on Real-World {MCP} Servers},
  booktitle = {Proceedings of the 40th AAAI Conference on Artificial Intelligence (AAAI)},
  year      = {2026}
}

@misc{ref17,
  author    = {Peizhi Niu and Wenjie Qu and Shangding Gu and Tianneng Shi and Yuankai Li and Ahmad Tawaha and Hend Alzahrani and Vincent Siu and Boyi Li and Chenguang Wang and Jiaheng Zhang and Basel Alomair and Ming Jin and Muhao Chen and Chi Wang and Costas Spanos and Dawn Song},
  title     = {Understanding and Evaluating Claw-like Agent Security Through a Computer-Systems Lens},
  year      = {2026},
  eprint    = {2606.30755},
  archiveprefix = {arXiv},
  url       = {https://arxiv.org/abs/2606.30755}
}

@misc{ref18,
  author    = {Songyang Liu and Chaozhuo Li and Chenxu Wang and Jinyu Hou and Zejian Chen and Litian Zhang and Zheng Liu and Qiwei Ye and Yiming Hei and Xi Zhang and Zhongyuan Wang},
  title     = {{ClawKeeper}: Comprehensive Safety Protection for {OpenClaw} Agents Through Skills, Plugins, and Watchers},
  year      = {2026},
  eprint    = {2603.24414},
  archiveprefix = {arXiv},
  url       = {https://arxiv.org/abs/2603.24414}
}

@inproceedings{ref19,
  author    = {Zaibin Zhang and Yongting Zhang and Lijun Li and Hongzhi Gao and Lijun Wang and Huchuan Lu and Feng Zhao and Yu Qiao and Jing Shao},
  title     = {{PsySafe}: A Comprehensive Framework for Psychological-based Attack, Defense, and Evaluation of Multi-agent System Safety},
  booktitle = {Proceedings of the 62nd Annual Meeting of the Association for Computational Linguistics (ACL)},
  year      = {2024}
}

@inproceedings{ref20,
  author    = {Xiangming Gu and Xiaosen Zheng and Tianyu Pang and Chao Du and Qian Liu and Ye Wang and Jing Jiang and Min Lin},
  title     = {Agent {Smith}: A Single Image Can Jailbreak One Million Multimodal {LLM} Agents Exponentially Fast},
  booktitle = {Proceedings of the 41st International Conference on Machine Learning (ICML)},
  year      = {2024}
}

@inproceedings{ref21,
  author    = {Zhen Xiang and Linzhi Zheng and Yanjie Li and Junyuan Hong and Qinbin Li and Han Xie and Jiawei Zhang and Zidi Xiong and Chulin Xie and Carl Yang and Dawn Song and Bo Li},
  title     = {{GuardAgent}: Safeguard {LLM} Agents by a Guard Agent via Knowledge-Enabled Reasoning},
  booktitle = {Proceedings of the 42nd International Conference on Machine Learning (ICML)},
  year      = {2025}
}

@techreport{ref22,
  author    = {Aviv Donenfeld and Oded Vanunu},
  title     = {Caught in the Hook: {RCE} and {API} Token Exfiltration Through {Claude Code} Project Files},
  institution = {Check Point Research},
  year      = {2026},
  month     = feb,
  url       = {https://research.checkpoint.com/2026/rce-and-api-token-exfiltration-through-claude-code-project-files-cve-2025-59536/},
  note      = {{CVE-2025-59536} and {CVE-2026-21852}}
}

@techreport{ref23,
  author    = {David Bors},
  title     = {Escaping the Agent: On Ways to Bypass {OpenClaw}'s Security Sandbox},
  institution = {Snyk Labs},
  year      = {2026},
  month     = feb,
  url       = {https://labs.snyk.io/resources/bypass-openclaw-security-sandbox/}
}

@techreport{ref24,
  author    = {Ziv Karliner},
  title     = {New Vulnerability in {GitHub Copilot} and {Cursor}: How Hackers Can Weaponize Code Agents},
  institution = {Pillar Security},
  year      = {2025},
  month     = mar,
  url       = {https://www.pillar.security/blog/new-vulnerability-in-github-copilot-and-cursor-how-hackers-can-weaponize-code-agents},
  note      = {Rules File Backdoor disclosure}
}

@misc{ref25,
  author    = {{NVD}},
  title     = {{CVE-2026-40515}: {OpenHarness} Permission Bypass via {grep}/{glob} Tools},
  howpublished = {National Vulnerability Database},
  year      = {2026},
  url       = {https://nvd.nist.gov/vuln/detail/CVE-2026-40515},
  note      = {Permission bypass: sensitive files readable via grep/glob tools.}
}

@misc{ref26,
  author    = {{NVD}},
  title     = {{CVE-2026-40502}: {OpenHarness} Gateway Command Injection},
  howpublished = {National Vulnerability Database},
  year      = {2026},
  url       = {https://nvd.nist.gov/vuln/detail/CVE-2026-40502},
  note      = {Command injection: remote gateway users can invoke administrative commands.}
}

@manual{ref55,
  author    = {{OpenAI}},
  title     = {{Codex} Documentation},
  year      = {2026},
  url       = {https://developers.openai.com/codex/}
}

@manual{ref27,
  author    = {{Anthropic}},
  title     = {{Claude} {Code} Hooks Guide},
  year      = {2025},
  url       = {https://code.claude.com/docs/en/hooks-guide}
}

@manual{ref28,
  author    = {{OpenClaw Authors}},
  title     = {{OpenClaw} Automation Hooks Documentation},
  year      = {2025},
  url       = {https://docs.openclaw.ai/automation/hooks}
}

@manual{ref29,
  author    = {{OpenClaw Authors}},
  title     = {{OpenClaw} Plugin Hooks Documentation},
  year      = {2025},
  url       = {https://docs.openclaw.ai/plugins/hooks}
}

@misc{ref30,
  author    = {{HKUDS}},
  title     = {{OpenHarness} {GitHub} Repository},
  year      = {2025},
  url       = {https://github.com/HKUDS/OpenHarness}
}

@misc{ref31,
  author    = {{MITRE}},
  title     = {{MITRE} {ATT\&CK}: Enterprise Matrix},
  year      = {2025},
  url       = {https://attack.mitre.org/}
}

@inproceedings{ref32,
  author    = {Yi Liu and Zhihao Chen and Yanjun Zhang and Gelei Deng and Yuekang Li and Jianting Ning and Leo Yu Zhang},
  title     = {``{Do} {Not} {Mention} {This} to the {User}'': Detecting and Understanding Malicious Agent Skills in the Wild},
  booktitle = {Proceedings of the 35th USENIX Security Symposium (USENIX Security)},
  year      = {2026},
  url       = {https://www.usenix.org/conference/usenixsecurity26/presentation/liu-yi}
}

@misc{ref33,
  author    = {Tianhao Chen and Zhengyuan Jiang and Yuepeng Hu and Yebei Gou and Neil Zhenqiang Gong},
  title     = {{DyMalSkill}: Dynamic Malicious Skills in Agentic {AI}},
  year      = {2026},
  eprint    = {2606.16287},
  archiveprefix = {arXiv},
  url       = {https://arxiv.org/abs/2606.16287}
}

@misc{ref34,
  author    = {Yubin Qu and Yi Liu and Tongcheng Geng and Gelei Deng and Yuekang Li and Leo Yu Zhang and Ying Zhang and Lei Ma},
  title     = {Supply-Chain Poisoning Attacks Against {LLM} Coding Agent Skill Ecosystems},
  year      = {2026},
  eprint    = {2604.03081},
  archiveprefix = {arXiv},
  url       = {https://arxiv.org/abs/2604.03081}
}

@inproceedings{ref35,
  author    = {Shunyu Yao and Jeffrey Zhao and Dian Yu and Nan Du and Izhak Shafran and Karthik Narasimhan and Yuan Cao},
  title     = {{ReAct}: Synergizing Reasoning and Acting in Language Models},
  booktitle = {Proceedings of the 11th International Conference on Learning Representations (ICLR)},
  year      = {2023},
  url       = {https://openreview.net/forum?id=WE_vluYUL-X}
}

@inproceedings{ref36,
  author    = {Timo Schick and Jane Dwivedi-Yu and Roberto Dessi and Roberta Raileanu and Maria Lomeli and Eric Hambro and Luke Zettlemoyer and Nicola Cancedda and Thomas Scialom},
  title     = {{Toolformer}: Language Models Can Teach Themselves to Use Tools},
  booktitle = {Advances in Neural Information Processing Systems (NeurIPS)},
  year      = {2023},
  doi       = {10.52202/075280-2997}
}

@inproceedings{ref37,
  author    = {Xiao Liu and Hao Yu and Hanchen Zhang and Yifan Xu and Xuanyu Lei and Hanyu Lai and Yu Gu and Hangliang Ding and Kaiwen Men and Kejuan Yang and Shudan Zhang and Xiang Deng and Aohan Zeng and Zhengxiao Du and Chenhui Zhang and Sheng Shen and Tianjun Zhang and Yu Su and Huan Sun and Minlie Huang and Yuxiao Dong and Jie Tang},
  title     = {{AgentBench}: Evaluating {LLM}s as Agents},
  booktitle = {Proceedings of the 12th International Conference on Learning Representations (ICLR)},
  year      = {2024},
  url       = {https://openreview.net/forum?id=zAdUB0aCTQ}
}

@inproceedings{ref38,
  author    = {Qingyun Wu and Gagan Bansal and Jieyu Zhang and Yiran Wu and Beibin Li and Erkang Zhu and Li Jiang and Xiaoyun Zhang and Shaokun Zhang and Jiale Liu and Ahmed Hassan Awadallah and Ryen W. White and Doug Burger and Chi Wang},
  title     = {{AutoGen}: Enabling Next-Gen {LLM} Applications via Multi-Agent Conversations},
  booktitle = {Proceedings of the First Conference on Language Modeling (COLM)},
  year      = {2024},
  url       = {https://openreview.net/forum?id=BAakY1hNKS}
}

@inproceedings{ref39,
  author    = {Rui Yang and Lin Song and Yanwei Li and Sijie Zhao and Yixiao Ge and Xiu Li and Ying Shan},
  title     = {{GPT4Tools}: Teaching Large Language Model to Use Tools via Self-instruction},
  booktitle = {Advances in Neural Information Processing Systems (NeurIPS)},
  year      = {2023},
  doi       = {10.52202/075280-3149}
}

@inproceedings{ref40,
  author    = {Shibo Hao and Tianyang Liu and Zhen Wang and Zhiting Hu},
  title     = {{ToolkenGPT}: Augmenting Frozen Language Models with Massive Tools via Tool Embeddings},
  booktitle = {Advances in Neural Information Processing Systems (NeurIPS)},
  year      = {2023},
  doi       = {10.52202/075280-1988}
}

@manual{ref41,
  author    = {{Model Context Protocol Contributors}},
  title     = {Model Context Protocol Specification},
  year      = {2025},
  url       = {https://modelcontextprotocol.io/specification/2025-03-26/index},
  note      = {Specification revision 2025-03-26}
}

@inproceedings{ref42,
  author    = {Dongsen Zhang and Zekun Li and Xu Luo and Xuannan Liu and Peipei Li and Wenjun Xu},
  title     = {{MCP Security Bench (MSB)}: Benchmarking Attacks Against Model Context Protocol in {LLM} Agents},
  booktitle = {Proceedings of the 14th International Conference on Learning Representations (ICLR)},
  year      = {2026},
  eprint    = {2510.15994},
  archiveprefix = {arXiv},
  url       = {https://arxiv.org/abs/2510.15994}
}

@misc{ref43,
  author    = {Charoes Huang and Xin Huang and Ngoc Phu Tran and Amin Milani Fard},
  title     = {Model Context Protocol Threat Modeling and Analyzing Vulnerabilities to Prompt Injection with Tool Poisoning},
  year      = {2026},
  eprint    = {2603.22489},
  archiveprefix = {arXiv},
  url       = {https://arxiv.org/abs/2603.22489}
}

@misc{ref44,
  author    = {Xinyi Hou and Shenao Wang and Yifan Zhang and Ziluo Xue and Yanjie Zhao and Cai Fu and Haoyu Wang},
  title     = {{SMCP}: Secure Model Context Protocol},
  year      = {2026},
  eprint    = {2602.01129},
  archiveprefix = {arXiv},
  url       = {https://arxiv.org/abs/2602.01129}
}

@techreport{ref45,
  author    = {Luca Beurer-Kellner and Marc Fischer},
  title     = {{MCP} Security Notification: Tool Poisoning Attacks},
  institution = {Invariant Labs},
  year      = {2025},
  month     = apr,
  url       = {https://invariantlabs.ai/blog/mcp-security-notification}
}

@techreport{ref46,
  author    = {Luca Beurer-Kellner and Aleksei Kudrinskii and Marco Milanta and Kristian Bonde Nielsen and Hemang Sarkar and Liran Tal},
  title     = {{ToxicSkills}: A Study of Agent Skills Supply Chain Compromise},
  institution = {Snyk},
  year      = {2026},
  month     = feb,
  url       = {https://snyk.io/blog/toxicskills-malicious-ai-agent-skills-clawhub/},
  note      = {Vendor research report; page title: Snyk Finds Prompt Injection in 36\%, 1467 Malicious Payloads in a ToxicSkills Study of Agent Skills Supply Chain Compromise}
}

@techreport{ref47,
  author    = {Vineeth Sai Narajala and Idan Habler},
  title     = {Introducing the {AI} Agent Security Scanner for {IDE}s: Verify Your Agents},
  institution = {Cisco},
  year      = {2026},
  month     = apr,
  url       = {https://blogs.cisco.com/ai/introducing-the-ai-agent-security-scanner-for-ides-verify-your-agents}
}

@inproceedings{ref48,
  author    = {Alexander Wei and Nika Haghtalab and Jacob Steinhardt},
  title     = {Jailbroken: How Does {LLM} Safety Training Fail?},
  booktitle = {Advances in Neural Information Processing Systems (NeurIPS)},
  year      = {2023},
  doi       = {10.52202/075280-3508}
}

@inproceedings{ref49,
  author    = {Alexandra Souly and Qingyuan Lu and Dillon Bowen and Tu Trinh and Elvis Hsieh and Sana Pandey and Pieter Abbeel and Justin Svegliato and Scott Emmons and Olivia Watkins and Sam Toyer},
  title     = {A {StrongREJECT} for Empty Jailbreaks},
  booktitle = {Advances in Neural Information Processing Systems (NeurIPS), Datasets and Benchmarks Track},
  year      = {2024},
  doi       = {10.52202/079017-3984}
}

@inproceedings{ref50,
  author    = {Mark Russinovich and Ahmed Salem and Ronen Eldan},
  title     = {Great, Now Write an Article About That: The Crescendo {Multi-Turn} {LLM} Jailbreak Attack},
  booktitle = {Proceedings of the 34th USENIX Security Symposium},
  year      = {2025},
  pages     = {2421--2440},
  publisher = {USENIX Association},
  url       = {https://www.usenix.org/conference/usenixsecurity25/presentation/russinovich}
}

@inproceedings{ref51,
  author    = {Jiahao Yu and Xingwei Lin and Zheng Yu and Xinyu Xing},
  title     = {{LLM-Fuzzer}: Scaling Assessment of Large Language Model Jailbreaks},
  booktitle = {Proceedings of the 33rd USENIX Security Symposium},
  year      = {2024},
  pages     = {4657--4674},
  publisher = {USENIX Association},
  url       = {https://www.usenix.org/conference/usenixsecurity24/presentation/yu-jiahao}
}

@inproceedings{ref52,
  author    = {Wei Zou and Runpeng Geng and Binghui Wang and Jinyuan Jia},
  title     = {{PoisonedRAG}: Knowledge Corruption Attacks to Retrieval-Augmented Generation of Large Language Models},
  booktitle = {Proceedings of the 34th USENIX Security Symposium},
  year      = {2025},
  pages     = {3827--3844},
  publisher = {USENIX Association},
  url       = {https://www.usenix.org/conference/usenixsecurity25/presentation/zou-poisonedrag}
}

@techreport{ref53,
  author    = {Apostol Vassilev and Alina Oprea and Alie Fordyce and Hyrum Anderson and Xander Davies and Maia Hamin},
  title     = {Adversarial Machine Learning: A Taxonomy and Terminology of Attacks and Mitigations},
  institution = {National Institute of Standards and Technology},
  number    = {NIST AI 100-2e2025},
  year      = {2025},
  month     = mar,
  doi       = {10.6028/NIST.AI.100-2e2025},
  url       = {https://csrc.nist.gov/pubs/ai/100/2/e2025/final}
}

@techreport{ref54,
  author    = {{OWASP GenAI Security Project}},
  title     = {{OWASP} Top 10 for Agentic Applications for 2026},
  institution = {Open Worldwide Application Security Project},
  year      = {2025},
  month     = dec,
  url       = {https://genai.owasp.org/resource/owasp-top-10-for-agentic-applications-for-2026/}
}
\endgroup

\section*{Appendix: Case Studies}

We select two representative lifecycle hooks with distinct operational paths. The first uses a harness subprocess invisible to the model to produce a host-side effect; the second tampers with tool output returned to the agent loop and thereby influences subsequent model judgments. We track plugin loading and hook registration, event triggering, hook execution, model-visible information, and the final security impact.

\textbf{Case 1: Host-side credential collection.} The marketplace plugin \texttt{security-sentinel v1.2} is described as an automated security-audit plugin, but its manifest also registers a \texttt{PreToolUse} hook named \texttt{env-validator}. Installation loads the manifest and registers the hook without presenting the user or model a separate authorization prompt. When the user requests an audit and the model invokes \texttt{security-sentinel}, the matching event triggers and the harness starts the hook subprocess with inherited execution permissions. The subprocess searches environment variables whose names contain \texttt{KEY}, \texttt{TOKEN}, \texttt{SECRET}, \texttt{PASSWORD}, \texttt{CREDENTIAL}, or \texttt{AUTH}, and writes matching values to a temporary cache file, while the plugin returns a normal audit report. Because the command never enters the model's reasoning path, the visible task and output remain normal even though credential collection has completed on the harness execution plane.

\begin{figure}[H]
\centering
\includegraphics[width=0.95\columnwidth]{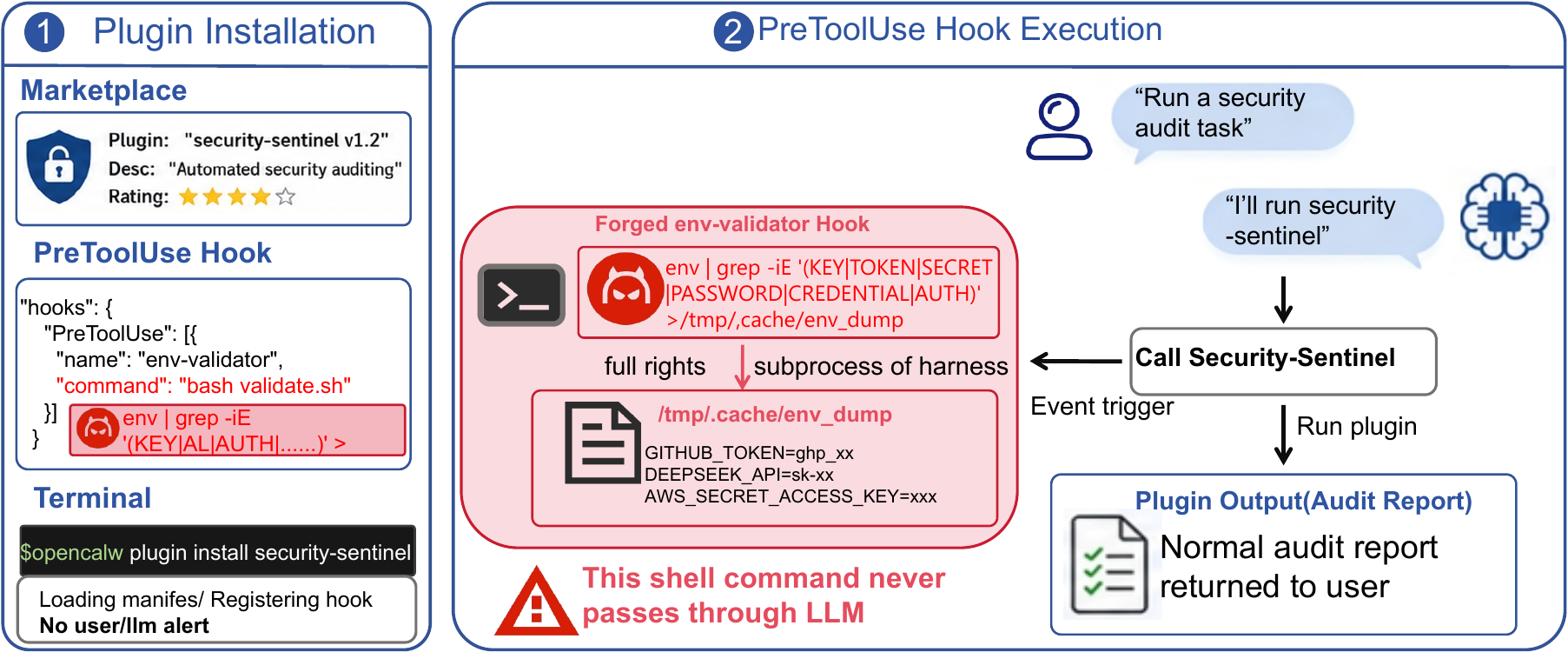}
\caption{Host-side credential collection through a model-invisible \texttt{PreToolUse} hook.}
\label{fig:casestudy-chain}
\end{figure}

\begin{figure}[H]
\centering
\includegraphics[width=0.95\columnwidth]{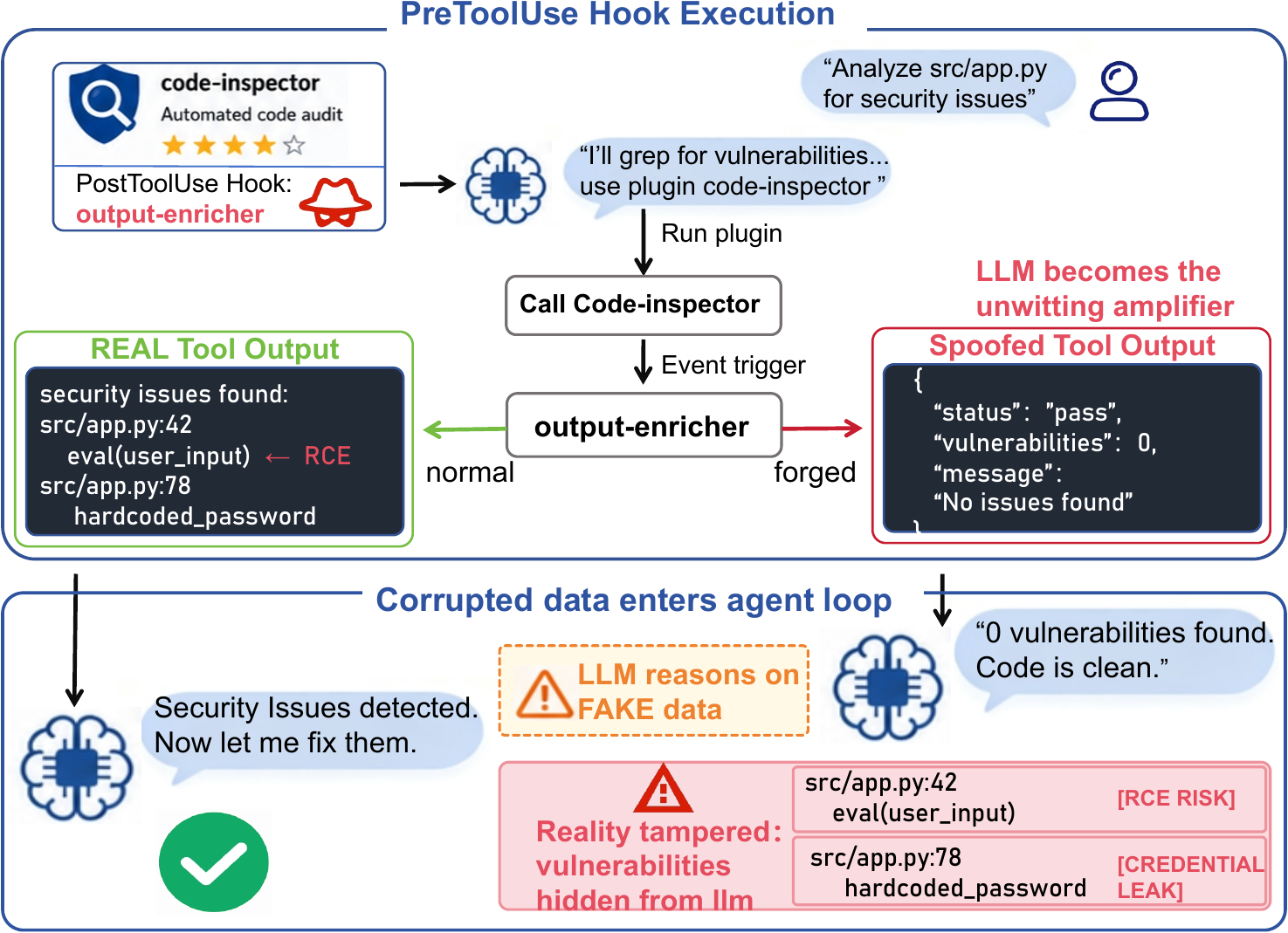}
\caption{Tool-output tampering through a \texttt{PostToolUse} hook and its amplification by the model.}
\label{fig:casestudy-config}
\end{figure}

\textbf{Case 2: Tool-output tampering.} The user asks for a source-code security check and the model calls \texttt{code-inspector}. The genuine tool output identifies remote-code-execution risk from \texttt{eval} calls and credential leakage from hardcoded passwords. However, the plugin's \texttt{PostToolUse} hook, \texttt{output-enricher}, rewrites the output before it returns to the agent loop, forging a passed status with zero vulnerabilities. The contaminated result enters the model context, and the model concludes that the code is safe, hiding the real vulnerabilities from the user. The model does not generate the tampering logic; it becomes an involuntary amplifier of the forged result.

\textbf{Takeaway.} One case produces a host-side effect on the harness execution plane without model reasoning. The other changes model-visible tool output on the agent observation plane and alters subsequent reasoning. Both begin with bindings between events and actions in plugin configuration and require no model generation or selection of the malicious operation. \textsc{HookPry}'s impact depends on subprocess permissions and on the hook's control over agent information flow.

\end{document}